\documentclass[12pt,a4paper]{article}
\pdfoutput=1

\usepackage{jheppub}
\usepackage[usenames,dvipsnames]{xcolor}
\definecolor{MathematicaBlue}{rgb}{0.368417, 0.506779, 0.709798}
\definecolor{MathematicaRed}{rgb}{0.922526, 0.385626, 0.209179}
\usepackage{hyperref}
\hypersetup{
colorlinks=true,
urlcolor=Maroon,
linkcolor=RoyalBlue,
citecolor=Maroon,
pdftitle={On Unitarity of Tree-Level String Amplitudes},
pdfauthor={},
pdfdisplaydoctitle=true,
pdfstartview=FitH,
linktocpage=true
}

\usepackage{amssymb} 
\usepackage{amsmath}
\usepackage{mathtools}
\usepackage{amsfonts}    
\usepackage{dsfont}
\usepackage{pdfpages}
\usepackage{verbatim}
\usepackage{tensor}
\usepackage{mathrsfs}
\usepackage[mathscr]{euscript}
\usepackage{tikz}
\usetikzlibrary{decorations.pathmorphing}
\usepackage{tcolorbox}
\usepackage{nccmath}
\usepackage[export]{adjustbox}

\usepackage[smalltableaux]{ytableau}

\newcommand{\ba}{\begin{align}}

\newcommand{\be}{\begin{equation}}
\newcommand{\ee}{\end{equation}}
\def\bd{\begin{tikzpicture}}
\def\ed{\end{tikzpicture}}

\renewcommand{\d}{\mathrm{d}}
\renewcommand{\geq}{\geqslant}
\renewcommand{\ge}{\geqslant}
\renewcommand{\leq}{\leqslant}
\renewcommand{\le}{\leqslant}
\DeclareMathOperator{\Res}{Res}
\newcommand{\nn}{\nonumber}

\newcommand{\I}{\mathrm{I}}
\newcommand{\II}{\mathrm{II}}
\newcommand{\open}{\mathrm{open}}
\newcommand{\closed}{\mathrm{closed}}
\newcommand{\bosonic}{\mathrm{bos}}
\newcommand{\het}{\mathrm{het}}

\allowdisplaybreaks[1]

\begin{document}

\title{On Unitarity of Tree-Level String Amplitudes}

\author[1]{Nima Arkani-Hamed,}\emailAdd{arkani@ias.edu}
\author[1]{Lorenz Eberhardt,}\emailAdd{elorenz@ias.edu}
\author[2,3]{Yu-tin Huang,}\emailAdd{yutinyt@gmail.com}
\author[1]{Sebastian Mizera}\emailAdd{smizera@ias.edu}

\affiliation[1]{Institute for Advanced Study, Einstein Drive, Princeton, NJ 08540, USA}
\affiliation[2]{Department of Physics and Center for Theoretical Physics, National Taiwan University, Taipei 10617, Taiwan}
\affiliation[3]{Physics Division, National Center for Theoretical Sciences, Taipei 10617, Taiwan}

\abstract{Four-particle tree-level scattering amplitudes in string theory are magically consistent with unitarity, reflected in the non-trivial fact that beneath the critical dimension, the residues of the amplitudes on massive poles can be expanded in partial waves with all positive coefficients. While this follows (rather indirectly) from the no-ghost theorem, the simplicity of the statement and its fundamental importance for the physical consistency of string theory begs for a more direct and elementary understanding. In this note we take a step in this direction by presenting a new expression for the partial wave coefficients of string amplitudes, given by surprisingly simple double/triple contour integrals for open/closed strings. This representation allows us to directly prove unitarity of all superstring theories in $D \leq 6$ spacetime dimensions, and can also be used to determine various asymptotics of the partial waves at large mass levels.}

\maketitle
\setcounter{page}{2}

\makeatletter
\g@addto@macro\bfseries{\boldmath}
\makeatother

\section{Introduction} \label{sec:intro}
The four-point tree-level amplitude in string theory is a famously miraculous object \cite{Veneziano:1968yb,Virasoro:1969me,Shapiro:1970gy}. Consider for instance the color-ordered four-tachyon amplitude of open bosonic strings, where (with $\alpha'=1$)
\begin{equation}
A^{\open,\bosonic }(s,t) =\frac{\Gamma({-}s{-}1)\Gamma({-}t{-}1)}{\Gamma(2{+}u)}\,.
\end{equation}
The above expression satisfies all the stringent consistency constraints on amplitudes in one shot. It is consistent with causality, by being appropriately bounded by a power of $s$ for large centre-of-mass energy $s$, working in the Regge limit at fixed momentum transfer $t$. It is famously UV complete, being exponentially soft for physical Lorentzian kinematics ($stu>0$ with $s{+}t{+}u=-4$), at large energy and fixed angle.

But by far the most miraculous property of the amplitudes is the way they are consistent with unitarity.  Consider the residue of $A^{\open,\bosonic}(s,t)$ as we approach a pole at $s \to n$. We have $A^{\open,\bosonic}(s,t) \to \frac{1}{s {-} n} \times R^{\open,\bosonic}_n(t)$ with $R^{\open,\bosonic}_n(t) = \frac{(2{+}t)(3{+}t) \cdots(n{+}2 {+} t)}{(n{+}1)!}$. As usual, we can express this result in terms of the scattering angle $x = \cos \theta$ with $t = (s{+}4)(x{-}1)/2 = (n{+}4)(x{-}1)/2$, so that the ``residue polynomial" $R^{\open,\bosonic}_n(x)$ is
\be
R_n^{{\open,\bosonic}}(x) = \frac{1}{(n{+}1)!}\left(\frac{n{+}4}{2}\right)^{\!\!n{+}1}\prod_{i=1}^{n{+}1}\left(x {-} \frac{n {-} 2 i{+} 2}{n{+}4}\right)\,.
\ee
Unitarity demands that $R^{\open,\bosonic}_n(x)$ can be expanded in Gegenbauer polynomials $G_j^{(D)}(x)$ with all positive or zero coefficients: $R^{\open,\bosonic}_n(x) = \sum_j B_{n,j}^{D}\, G_j^{(D)}(x)$ with  $B_{n,j}^{D} \geq 0$ and spacetime dimension $D$. The positivity of the $B_{n,j}^D$'s follows from the fact that they are the mod-square of three-particle amplitude couplings to the internal states of spin-$j$, or equivalently, represent the cross-section for the resonant production of the intermediate spin-$j$ states. Here the first interesting level is $n=1$, where we find 
\be \frac{8}{25}R_{n=1}^{\open,\bosonic}(x) = \left(x{-}\frac{1}{5}\right)\left(x{+}\frac{1}{5}\right)  = \left(x^2 {-} \frac{1}{D{-}1}\right) + \left(\frac{1}{D{-}1}{-}\frac{1}{25}\right)\,.
\ee 
The spin-2 Gegenbauer polynomial in $D$ dimensions is proportional to $(x^2 {-} \frac{1}{D{-}1})$, so this shows that we are exchanging a spin-2 state with a positive norm. But the remaining constant term indicates the presence of a spin-0 state with the norm $\frac{1}{D{-}1}{-}\frac{1}{25}$. This gives a negative-norm state for $D > 26$ spacetime dimensions. In this way, the four-point tree-level amplitude---whose form is completely independent of spacetime dimensionality---knows about the critical dimension, through unitarity.

The story is much more interesting for the scattering of spinning states. For the four-gluon amplitude in type-I superstrings,  
\begin{equation}
   {\cal A}^{\I}(s,t) = \mathcal{F}^4 A^{\I}(s,t)  = \mathcal{F}^4 \frac{\Gamma({-}s) \Gamma({-}t)}{\Gamma(1{+}u)} \label{eq:four gluon amplitude}\,.
\end{equation}
Here, $\mathcal{F}^4$ is a famous permutation-invariant polynomial in the field strengths that appears in the tree-level Yang--Mills amplitude: $A^{\rm YM}=\frac{ \mathcal{F}^4}{st}$. Unitarity at the first massive level is already non-trivial: as the residue is simply $\mathcal{F}^4$ itself, the permutation-invariant polynomial must be written as a positive sum of $s$-channel exchanges. With  $x=p_1{-} p_2$ and $y=p_3{-}p_4$, we express
\begin{equation}
\mathcal{F}^4(\epsilon_i, x,y)=\langle x, \epsilon_1, \epsilon_2| \mathcal{O}| y,\epsilon_3,\epsilon_4\rangle\,,
\end{equation}
where $| x,\epsilon_1,\epsilon_2\rangle$ represents the ``Hilbert space" of states with $x, \epsilon_1, \epsilon_2$. From these states we can build irreps of SO($D{-}1$), on which  $\mathcal{O}$ can be expanded. We find that in this form, $\mathcal{F}^4$ is uniquely given as a combination of spin-0, -2 and a 3 form:
\begin{equation}\label{F4Decomp}
\mathcal{O}=\left(\frac{9}{D{-}1}{-}1\right)\big| \cdot\big\rangle\big\langle \cdot\big|+\big|\,{\tiny \ytableausetup{centertableaux}\ydiagram{2}}\,\big\rangle\big\langle \,{\tiny \ytableausetup{centertableaux}\ydiagram{2}}\,\big|+\big|\,{\tiny \ytableausetup{centertableaux}\ydiagram{1,0+1,0+1}}\,\big\rangle\big\langle \,{\tiny \ytableausetup{centertableaux}\ydiagram{1,0+1,0+1}}\,\big| \,.
\end{equation}
Positivity of the expansion now requires $D \leq 10$. Note that this massive spectrum is exactly the bosonic content of 11-dimensional supergravity, as it must from the viewpoint of supersymmetry, i.e., massive irreps can be obtained from massless ones in one dimension higher. Thus, in a precise sense, the tree-level amplitude of Yang--Mills knows about the critical dimension of type-I superstring as well as supersymmetry! 

Let us also mention that the exchanged three fields correspond exactly to the field content expected from the worldsheet description of type-I strings. These fields correspond to the first massive level. Letting $\partial X^\mu$ and $\psi^\mu$ be the free bosons and fermions respectively, the unique two bosonic physical states in string theory at this mass-level read 
\be
    \big|\,{\tiny \ytableausetup{centertableaux}\ydiagram{2}}  \big\rangle \equiv \epsilon_{\mu \nu} \partial X^{\mu} \psi^{\nu}  e^{i p \cdot X}\ ,  \, \,\qquad 
    \big|\,{\tiny \ytableausetup{centertableaux}\ydiagram{1,0+1,0+1}}\,\big\rangle \equiv \epsilon_{\mu \nu \lambda}\psi^\mu \psi^\nu \psi^\lambda\, e^{i p \cdot X}\ ,
\ee
where $\epsilon_{\mu \nu}$ and $\epsilon_{\mu \nu \lambda}$ are the transverse traceless symmetric and antisymmetric polarization vectors respectively, i.e., $p^\mu \epsilon_{\mu \nu }=0$ and $p^\mu \epsilon_{\mu \nu \lambda}=0$. 
The scalar state is absent in ten-dimensional string theory, as predicted from the coefficient in \eqref{F4Decomp}. In the presence of an internal compactification manifold, one can generically write down one more scalar by choosing (the coefficients are chosen in order to ensure a traceless transverse polarization)
\be 
\epsilon_{\mu \nu}=\eta_{ab}+p_a p_b-\frac{D-1}{D-10} g_{ij}\ ,
\ee
where $a$, $b=0,1,\dots,D$ denote the flat directions and $i$, $j$ are the indices of the internal manifold with metric $g_{ij}$. We can identify this additional scalar with the scalar appearing in the expansion \eqref{F4Decomp} for $D<10$.

Going beyond the first level we have $ {\cal A}^{\I}(s,t) \to \frac{1}{s {-} n}\mathcal{F}^4 \times R^{\I}_n(t)$, where $R^{\I}_n(t)=\frac{(1{+}t)(2{+}t)\cdots(n{-}1{+}t)}{n!}$. Once again, in terms of the scattering angle it is given as:
\begin{equation}
    R_n^{\I}(x) =\frac{1}{n!} \left(\frac{n}{2}\right)^{n{-}1}\prod_{i=1}^{n{-}1} \left(x{-} \frac{n {-} 2i}{n}\right)\,. \label{eq:residue polynomial}
\end{equation}
Expanding on the Gegenbauer polynomials, $R^{\I}_n(x)=\sum_j B_{n,j}^{D}\, G_j^{(D)}(x)$, this can now be combined with the expansion in eq. (\ref{F4Decomp}) to yield a ``diagonalized" representation of the residue:
\begin{equation}
\mathcal{F}^4R^{\I}_n=\langle x, \epsilon_1, \epsilon_2| \mathcal{O}_n| y,\epsilon_3,\epsilon_4\rangle\,, 
\end{equation}
where 
\begin{equation}
\mathcal{O}_n=\sum_j B_{n,j}^{D}\left\{\left(\frac{9}{D{-}1}{-}1\right)\big| \cdot\otimes j\big\rangle\big\langle \cdot\otimes j\big|+\big|\,{\tiny \ytableausetup{centertableaux}\ydiagram{2}}\otimes j\big\rangle\big\langle \,{\tiny \ytableausetup{centertableaux}\ydiagram{2}}\otimes j\big|+\big|\,{\tiny \ytableausetup{centertableaux}\ydiagram{1,0+1,0+1}}\otimes j\big\rangle\big\langle \,{\tiny \ytableausetup{centertableaux}\ydiagram{1,0+1,0+1}}\otimes j\big|\right\} \,. 
\end{equation}
The diagonalized is in quotes, since the states are constructed from tensor products of either the spin-0, 2 or 3-form, with a symmetric spin-$j$ irrep, i.e., the states are not necessary orthogonal, see Section~\ref{sec:superstring} for details. Nevertheless, if the coefficients $B_{n,j}^D$ are positive, the operator ${\cal O}_n$ is manifestly positive, establishing unitarity. We already know that unitarity for the first massive level requires $D\leq 10$. Unitarity can be established if we can show that $B_{n,j}^{D}\geq0$ for $D\leq 10$. This is already non-trivial for level $n=3$, 
\begin{equation}
   \frac{8}{3} R^{\I}_{n=3}(x) = \left(x{-}\frac{1}{3}\right)\left(x{+} \frac{1}{3}\right) = \left(x^2{-} \frac{1}{D{-}1}\right) + \left(\frac{1}{D{-}1} {-} \frac{1}{9}\right)\,.
\end{equation}
Once again, we see that for $B_{3,0}^{D}\geq0$ requires  $D\leq 10$.

The remarkable claim is that this pattern holds for \emph{all} higher levels $n$: the residue polynomial can be expanded as 
\begin{equation}
R^{\I}_n(x) = \sum_j
B^D_{n,j} G_j^{(D)}(x), \qquad {\rm where} \qquad B^D_{n,j} \geq 0 \quad {\rm for} \quad D \leq 10\,.
\end{equation}
The coefficients $B^D_{n,j}$ for the first few levels are given in Table~\ref{tab:low levels}, and can be manually verified to be positive for $D \leq 10$. Note also the clear pattern of zeroes. These zeroes follow from the fact that
$R_n(x)$ is even in $x$ for odd $n$ and odd in $x$ for even $n$. Thus only even spins are exchanged for odd $n$ and vice-versa, so that $B_{n,j}^D$ vanishes when $n{+}j$ is even. 

\begin{table}
\begin{center}
\begin{tabular}{c|ccccc}
& $\medmath{j=0}$ & $\medmath{j=1}$ & $\medmath{j=2}$ & $\medmath{j=3}$ & $\medmath{j=4}$ \\
\hline
$\medmath{n=1}$ & $1$ & & & & \\
$\medmath{n=2}$ & 0 & $\frac{1}{2(D-3)}$ & & & \\
$\medmath{n=3}$ & $\frac{10-D}{24(D-1)}$ & 0 & $\frac{3}{4(D-1)(D-3)}$ & & \\
$\medmath{n=4}$ & 0 & $\frac{11-D}{12(D+1)(D-3)}$ & 0 & $\frac{2}{(D^2-1)(D-3)}$ & \\
$\medmath{n=5}$ & $\frac{9 D^2-250 D+1616}{1920 (D^2-1)}$ & 0 & $\frac{25 (12-D)}{96(D-1)(D^2-9)}$ & 0 & $\frac{125}{16 (D^2-1) (D^2-9)}$
\end{tabular}
\end{center}
\caption{\label{tab:low levels}Values of $B_{n,j}^D$ for the type-I amplitude with low $n$ and $j$. Note that all of them are non-negative for $D \leq 10$ and, e.g., $B_{3,0}^{D}$ indicates violation of unitarity for $D > 10$.} 
\end{table}

Note also that these fundamental positivity statements for open strings directly prove that the residue polynomials for closed strings also have a positive expansion. For example consider type-II superstring, heterotic Yang--Mills~\cite{Gross:1985rr}\footnote{Recall that the heterotic string contains the bosonic string in $R^{1,9}\times \Gamma^{16}$ as the left-moving part and the superstring in $R^{1,9}$ as the right moving part. The momenta on the bosonic side is now split into 10 continuous, and 16 discrete momenta associated with the torus $ \Gamma^{16}$. Here we consider four gluons whose color indices are root vectors, and arranging the discrete momenta $K_i$, with $K_i^2=1$, satisfying 
\begin{equation}
S=-(K_1{+}K_2)^2=0,\quad T=-(K_1{+}K_4)^2=0,\quad U=-(K_1{+}K_3)^2=-4\,,
\end{equation}
such that in the low-energy limit one directly lands on the Yang--Mills amplitude with $1234$ ordering.  } and closed tachyon amplitude are given by:
\begin{eqnarray}
\mathcal{A}^{\II}(s,t)&=&\mathcal{R}^4 A^{\II}(s,t) =\mathcal{R}^4 \frac{\Gamma({-}s) \Gamma({-}t) \Gamma({-}u)}{\Gamma(1 {+}s) \Gamma(1{+}t) \Gamma(1{+}u)},\nonumber\\
\mathcal{A}^{\rm Het}(s,t)&=&\mathcal{F}^4 A^{\rm Het}(s,t) =\mathcal{F}^4 \frac{\Gamma({-}s{-}1) \Gamma({-}t{-}1) \Gamma({-}u{+}3)}{\Gamma(1 {+}s) \Gamma(1{+}t) \Gamma(1{+}u)},\\
A^{{\closed, \bosonic}}(s,t)&=&-\frac{\Gamma({-}1{-}s)\Gamma({-}1{-}t)\Gamma({-}1{-}u)}{\Gamma(2{+}s)\Gamma(2{+}t)\Gamma(2{+}u)} \,,\nonumber
\end{eqnarray}
where $\mathcal{R}^4=(\mathcal{F}^4)^2$\,. The reason is simple: the residue for the type-II closed string is $\mathcal{R}^4 R_n^{\II}(x)=(\mathcal{F}^4 R^{\I}_n(x))^2$, and for the closed bosonic string is the square of the open bosonic string $R_n^{{\closed, \bosonic}}(x) = (R_n^{{\open, \bosonic}}(x))^2$, while the residue for the heterotic string, $\mathcal{F}^4 R_n^{\het}(x)$, can be written as:
\begin{equation}
R^{\het}_n(x) = \begin{dcases}
\frac{1}{8}(x{+}3)(x{+}5) &n=1,\\
\frac{1}{12}x(x{+}1)(x{+}2)(x{+}3) &n=2,\\
\frac{(1{+}x) (n^2 x (x{+}2) {+} n^2 {-} 4)(n(1{+}x){+}4)}{16 (n{-}2) (n{-}1) (n{+}1)} R_n^{\I}(x) R_{n{-}4}^{\open, \bosonic}(x) & n>2.
\end{dcases}
\end{equation}
For $n>2$, it is given by a product of type-I and open bosonic string residue with a completely positive function (a polynomial in $x$ with positive coefficients which is guaranteed to have a positive Gegenbauer expansion in any dimension).
But it is obvious that if $A(x)$ and $B(x)$ have a positive expansion on the Gegenbauers, so does the product $A(x) B(x)$. Indeed, this simply follows from $G_{j_1}(x) G_{j_2}(x)$ having a positive expansion, since spin-$j_1 \times$ spin-$j_2$ representations of $\mathrm{O}(D-1)$ obviously decompose as a positive sum of irreps of various spins. Thus, given that $R_n(x)$ has a positive expansion for the open strings, the products associated with the closed strings also have positive expansions. 
We should also notice that unitarity in higher dimensions immediately implies unitarity in lower dimensions for essentially the same reason. $G_j^{(D)}(x)$ decomposes into a positive sum of $G_{j'}^{(D')}(x)$ for $D' < D$ according to the branching rules of $\mathrm{O}(D-1)$ into $\mathrm{O}(D'-1)$. This is also physically obvious since we may always compactify a unitary theory in $D$ spacetime dimensions to obtain a unitary theory in $D' < D$ dimensions.

The Gegenbauer positivity of $R_n(x)$ is a strikingly elementary statement. There is an indirect proof of this positivity, via the famous no-ghost theorem \cite{Goddard:1972iy}; a general discussion of unitarity in closed string perturbation theory is given in \cite{Aoki:1990yn}. On-shell perspective on unitarity was previously explored in the context of high-energy expansion of partial waves \cite{Soldate:1986mk} and Hankel-matrix constraints on the Wilson coefficients \cite{Green:2019tpt}. There is also an interesting sort of converse to this statement, where the assumption of positivity for the Gegenbauer expansion of four-particle amplitudes, for {\it any} theory with an infinite tower of massive states, implies that the amplitude asymptotes to the string amplitudes in the (unphysical) high-energy regime where $s,t$ are both large and negative \cite{Caron-Huot:2016icg}. 

But the basic statement of positivity is so simple, and of such fundamental importance to the consistency of string theory, that it cries out for a more elementary and direct understanding on the level of the amplitudes. 
Indeed, some extreme cases are easy to understand. Most trivially, $R^{\I}_n(x) = \frac{1}{n!} (\frac{n}{2})^{n-1} x^{n-1} + \dots$; this tells us the highest spin at level $n$ is $j_{\max} = n{-}1$, and since the coefficient of $x^{n-1}$ is positive, also establishes the coefficient of the spin-$(n{-}1)$ Gegenbauer is positive. 

\begin{figure}[t]
\begin{center}
\includegraphics[scale=.5]{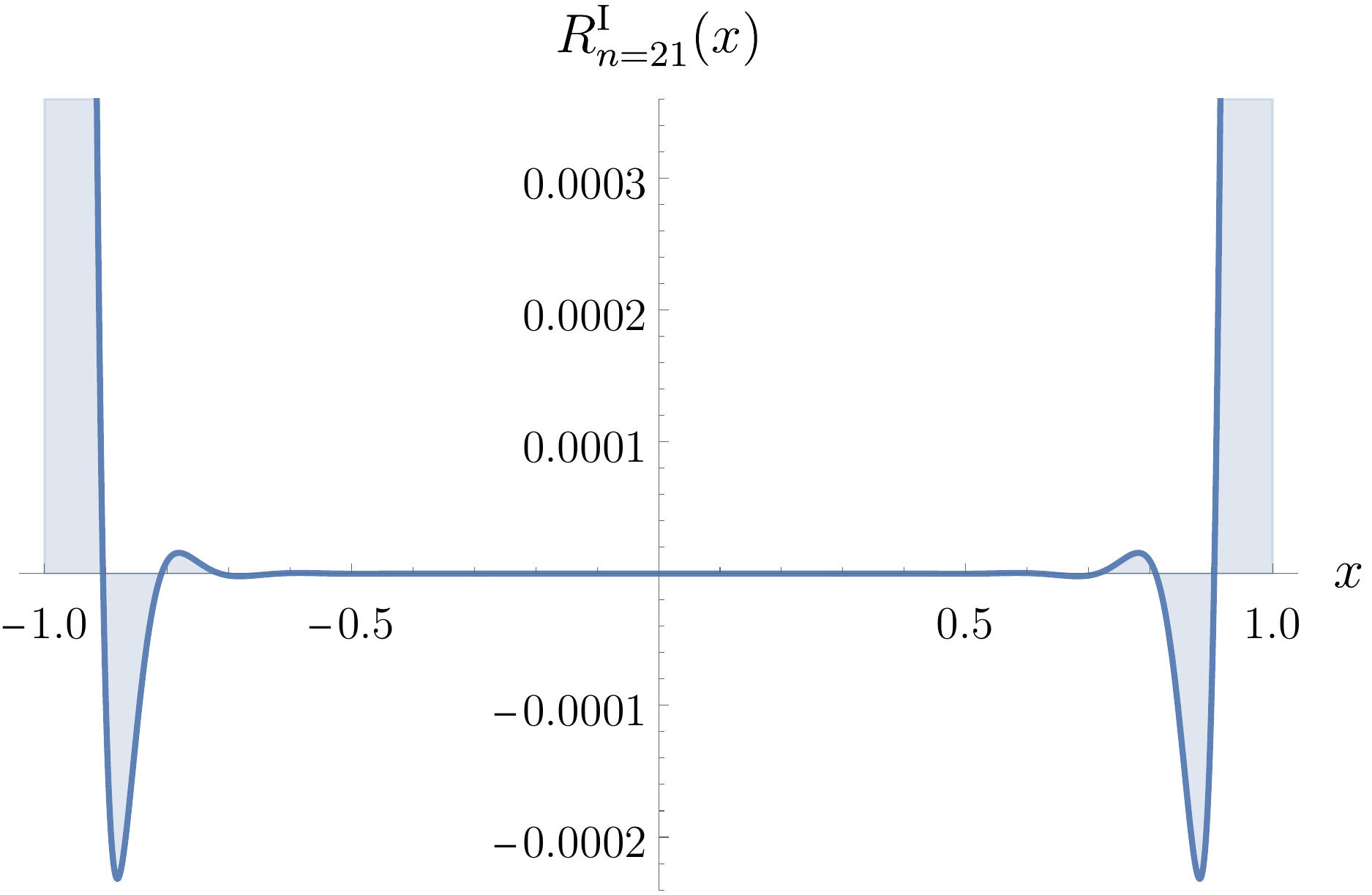}
\end{center}
\caption{\label{fig:R21}The residue coefficient $R_{n=21}^{\I}(x)$ as a function of the scattering angle $x= \cos \theta$.} 
\end{figure}

For general spin, the coefficient $B^D_{n,j}$ can be extracted using the orthogonality of Gegenbauer polynomials as 
\begin{equation}
    B^D_{n,j} \propto \int_{-1}^1 \d x\, (1 {-} x^2)^{\frac{D{-}4}{2}} G_j^{(D)}(x)\, R_n^{\I}(x)\, ,
    \end{equation}
where we suppressed the (positive) normalization factor coming from the normalization of the Gegenbauer polynomials.    
It is easy to see that for large $n$, $R^{\I}_n(x)$ is peaked around $x=\pm 1$, and is crushed away from the ends of the interval, as in Figure~\ref{fig:R21}.
This means that for fixed $j$ and large $n$, the integral for $B^D_{n,j}$ is dominated by the regions near $x = \pm 1$, where the contribution to $B^D_{n,j}$ is manifestly positive, so it is clear that for fixed $j$ and large enough $n$, $B^D_{n,j}$ is positive. 

\begin{figure}[t]
\begin{center}
\includegraphics[scale=.47]{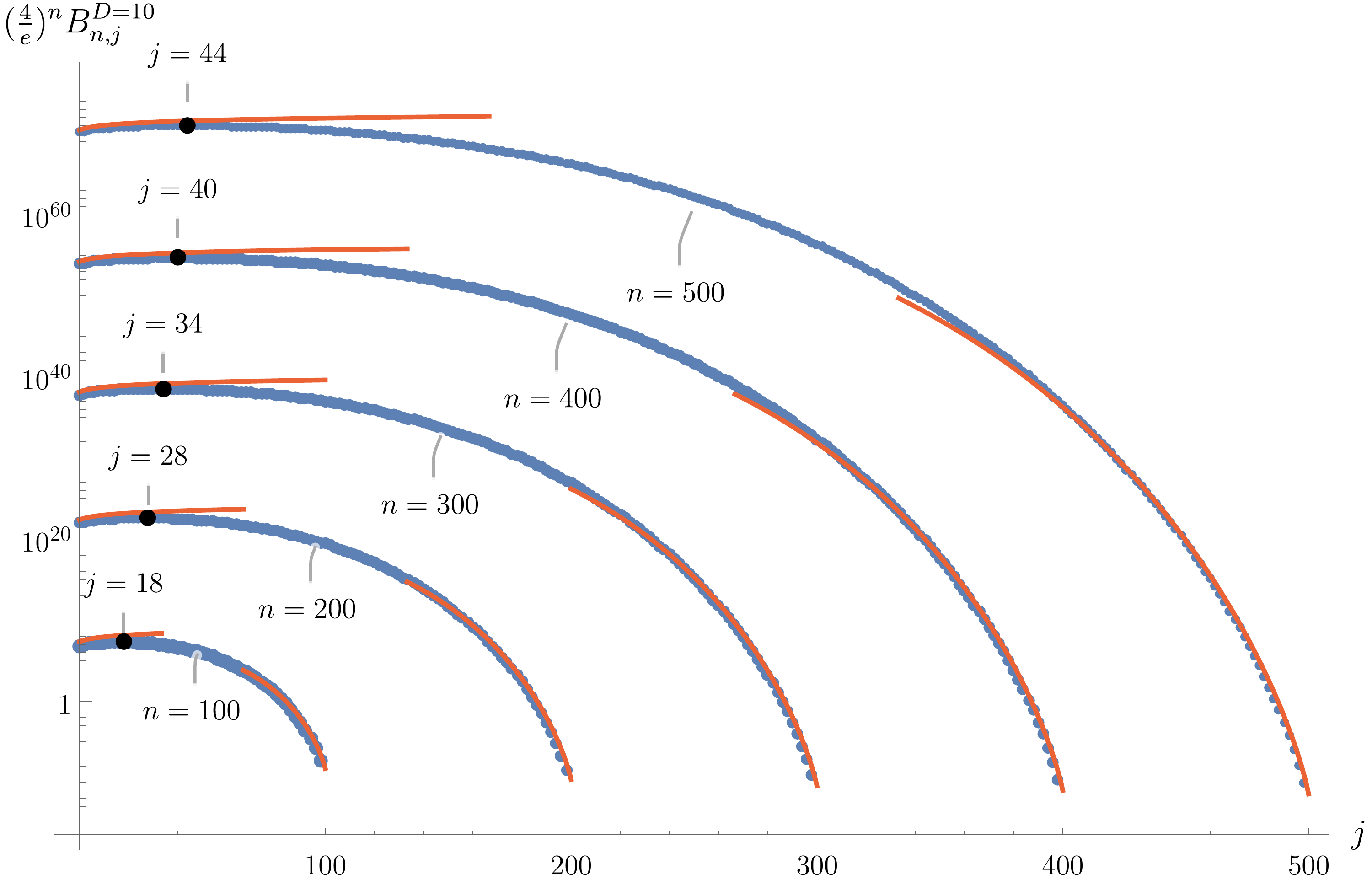}
\end{center}
\caption{\label{fig:coefficients}Coefficients $B^{D=10}_{n,j}$ for the type I superstring for five different values of $n$ plotted in blue against $j$ on a logarithmic scale. Only coefficients with $n{+}j$ odd are plotted and they have been rescaled by $(\tfrac{4}{e})^n$ to make the plot more readable. In each case, a peak is reached near $j \sim 2\sqrt{n}$, indicated with black dots, in qualitative agreement with the asymptotic formula for density of states derived in~\cite{Russo:1994ev}. The asymptotic formulas that are derived in Section~\ref{sec:asymptotics} for small-$j$ and small-$\Delta = n{-}j$ are both plotted in red.} 
\end{figure}

But for general $n$ and $j$, it is devilishly difficult to establish the positivity of $B^D_{n,j}$ by brute force. Almost all the ``obvious" ways of extracting $B^D_{n,j}$ are given as a sum of terms with alternating signs, with no evident reason for positivity. This is ironic, because it can be seen experimentally that the actual values of $B^D_{n,j}$ rise from a small positive value at $j=0$, to a maximum near $j \sim 2\sqrt{n}$, followed by a faster-than exponential decrease at larger $j$: it is hardest to prove $B^D_{n,j}$ is positive exactly where it has the largest positive magnitude!  This is illustrated in Figure~\ref{fig:coefficients}. The maximum at $j \sim 2\sqrt{n}$ is reminiscent of a similar peak in the distribution of spins contributing at a given mass level \cite{Curtright:1986di,Curtright:1986rr}.

In this note, we will revisit this basic question, by finding a remarkably simple new expression for the residues $B^D_{n,j}$ as a double-contour integral. Recall that $B^D_{n,j} = 0$ when $n{+}j$ is even; instead for $n{+}j$ odd we have, up to a positive constant\footnote{This formula as written only works for even $D$, but most of its consequences that we derive in this paper also work for odd $D$.}
\begin{tcolorbox}
\be 
B_{n,j}^D \propto \oint_{u=0} \frac{\mathrm{d}u}{2\pi i}\oint_{v=0} \frac{\mathrm{d}v}{2\pi i}\ \frac{(v-u)^j\, \mathrm{e}^{-a(u+v)}}{(uv)^{j+\frac{D-2}{2}} (\mathrm{e}^v-\mathrm{e}^u)^{n{+}2a}}\ . \label{eq:double contour formula}
\ee
\end{tcolorbox}
\noindent
Here $a=0$ and $1$ labels the type-I and bosonic open string amplitude respectively. We prove this formula in Section~\ref{sec:proof}.  
Using this expression, the positivity of $B^D_{n,j}$ in the type I case for all spacetime dimensions $D \leq 6$ follows easily, as shown in Section~\ref{sec:manifest unitarity}. For $6<D \leq 10$, the positivity is not as obviously manifest, though for any fixed spin $j$ can also be established for all $n$, with a finite amount of work. This is exemplified in Section~\ref{sec:unitarity in D=10}. Similar statements apply for the open bosonic string, where we can prove unitarity in the range $D \le 10$, while unitarity in the range $10<D \le 26$ can only be established for each fixed spin separately. The above expression also allows us to quantitatively analyze the asymptotic limits of $B^D_{n,j}$, for fixed $j$ at large $n$, and also for fixed Regge trajectories, taking $n$ large while keeping $\Delta = n{-}j$ fixed, see Section~\ref{sec:asymptotics}. These asymptotics are also indicated in Figure~\ref{fig:coefficients}.

Quite apart from its use in proving positivity and understanding the asymptotics, we find this double-contour integral representation of $B^D_{n,j}$ to be quite intriguing. While the string amplitudes themselves are clearly beautifully canonical objects, there was no a priori reason to expect any nice structure in the partial wave expansion of their residues! But such a structure {\it does} exist, and begs for a more direct understanding than the straightforward but not especially transparent derivation we will provide.  Indeed as we will see, the most striking feature of this expression --- its evident symmetry between the $u,v$ variables --- will not be apparent in our derivation at all, only popping out as a mysterious surprise at the end of the computation. 
There are analogous double-contour integral formulas for the open bosonic string, and triple-contour integrals for closed string residues. But we will leave the exploration of deeper origins for these formulae to future work; in this note we will content ourselves with deriving these unusual and interesting expressions, and highlighting their utility in understanding the fascinating positivity of the residues much more directly than previously possible.

\section{\label{sec:superstring}Unitarity of the superstring}
Let us begin with the gluon amplitude for the type-I superstring, given in eq.~(\ref{eq:four gluon amplitude}). The prefactor $\mathcal{F}^4$ is a permutation invariant polynomial of polarization vectors and momenta, which also appears as the numerator of the color-ordered $1234$ tree-level Yang--Mills amplitude as $A^{\mathrm{YM}} = \frac{\mathcal{F}^4}{ st}$, where 
\begin{equation}
    \mathcal{F}^4 = (F_{\mu \nu} F^{\mu \nu})^2 - 4 (F_{\mu \nu} F^{\nu \alpha} F_{\alpha \beta} F^{\beta \mu}).
\end{equation}
Here it is understood that we expand $F_{\mu \nu} = F_{1, \mu \nu} + \cdots + F_{4, \mu \nu}$, with $F_{i, \mu \nu} = p_{i, \mu} \epsilon_{i, \nu} - p_{i, \nu} \epsilon_{i, \mu}$, and we  extract only terms linear in each of the polarization vectors $\epsilon_{i, \mu}$.

We will find it most convenient to consider the scattering process in the center of mass frame, defined as 
\begin{equation}
\includegraphics[scale=.4,valign=c]{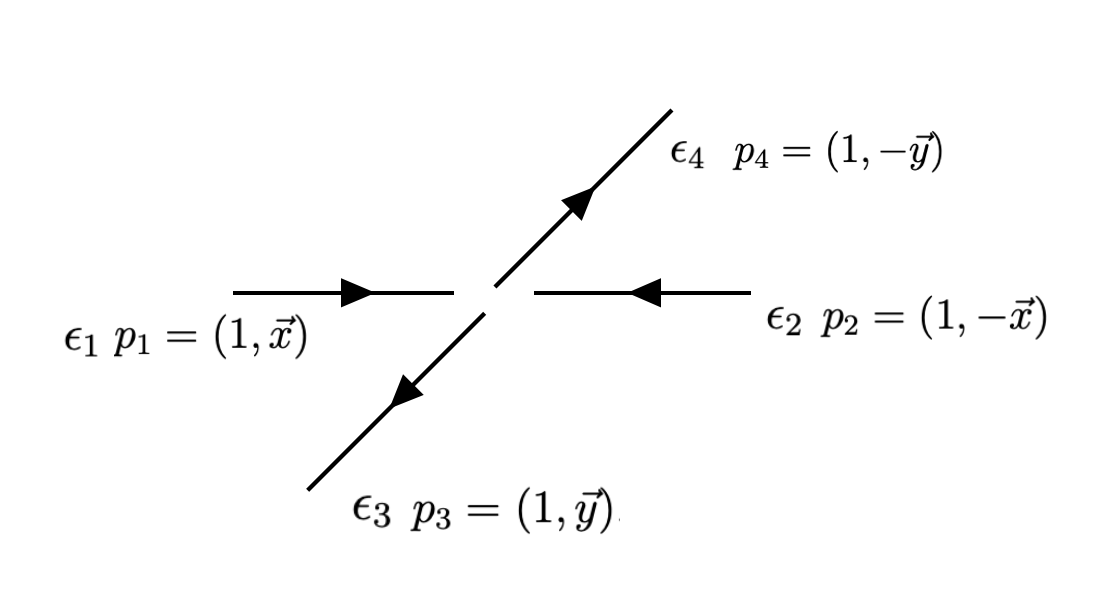}
\end{equation}
where $x, y$ are unit $(D{-}1)$-dimensional vectors, and the polarization vectors are chosen to be space-like $\epsilon_i=(0, E_i )$, satisfying $E_1\cdot x=E_2\cdot x=E_3\cdot y=E_4\cdot y=0$. A short computation then gives us $\mathcal{F}^4$ in terms of these center-of-mass frame variables:
\begin{eqnarray}\label{F4Form}
&&\mathcal{F}^4(E_i,x,y)=(E_1\cdot E_2)(E_3\cdot E_4)\left[(x\cdot y)^2-1\right]\\
&+&(E_2\cdot E_3)(E_1\cdot E_4)2\left[1{+}(x\cdot y)\right]+(E_1\cdot E_3)(E_2\cdot E_4)2\left[1{-}(x\cdot y)\right]\nonumber\\
&+&2\left[(E_1\cdot E_2)(E_3\cdot x)(E_4\cdot x){+}(E_3\cdot E_4)(E_1\cdot y)(E_2\cdot y){-}(E_2\cdot E_3)(E_1\cdot y)(E_4\cdot x)\right.\nonumber\\
&&\left.{+}(E_1\cdot E_3)(E_2\cdot y)(E_4\cdot x){+}(E_2\cdot E_4)(E_1\cdot y)(E_3\cdot x){-}(E_1\cdot E_4)(E_2\cdot y)(E_3\cdot x)\right]\,.\nonumber
\end{eqnarray}
The $n$-th level residue can then be written as 
\begin{equation}
\mathcal{F}^4(E_i,x,y)\cdot R_n^{\I}(x\cdot y)\equiv E_{1}^aE_{2}^bE_{3}^cE_{4}^d (\mathcal{F}^4)_{a,b,c,d}(x, y) \cdot R_n^{\I}(x\cdot y)\,.
\end{equation}
We will now introduce a Hilbert space $|w,\,_{a\, b}\rangle$, where $w$ is a unit vector representing the direction of the in/out scattering states in the center-of-mass frame, and $a,b$ denote the spin degrees of freedom. These states are orthonormal $\langle w^\prime, a^\prime, b^\prime|w, a,b \rangle = \delta_{S^{D-2}}(w,w^\prime) \delta_{a^\prime a} \delta_{b^\prime b}$. This continuum of states transform as highly reducible representations of SO($D-1$), and we will build linear combinations of them by integrating over the the sphere with various weights, that transform more conveniently, as familiar from the introduction of spherical harmonics. 

Now the residue defines an operator $\mathbb{R}_n$,  by giving its matrix elements as:
\begin{equation}
\langle y,\,_{c\,d}| \mathbb{R}_n|x,\,_{a\,b}\rangle=(\mathcal{F}^4)_{a,b,c,d}(x, y) \cdot R_n^{\I}(x\cdot y)\,.
\end{equation} 
We would like to demonstrate that $\mathbb{R}_n$ is a positive operator, i.e., that for all the states in the Hilbert space $|\psi \rangle$, $\langle \psi | \mathbb{R}_n |\psi \rangle \geq 0$. We will do this by showing that the operator can be written in the form  are $\sum_i c_i |i\rangle\langle i|$ with $c_i>0$ and where $|i\rangle$'s are distinct but not necessarily orthogonal states. 

Let us start with $n=1$, where $R_1^{\I}(x\cdot y)=1$. The goal is then to find a positive operator $\mathbb{R}$ such that $\langle y,\,_{c\,d}| \mathbb{R}|x,\,_{a\,b}\rangle=(\mathcal{F}^4)_{a,b,c,d}$.  To this end, we will build SO($D{-}1$) irreps out of  $|w,\,_{a\, b}\rangle$. For example, for spin-$0$, $2$ and totally anti-symmetric 3-form we can construct,
\begin{eqnarray}
\big| \cdot\big\rangle&=& \int_w |w,\, _{k\, k}\rangle,\nonumber\\
\big| i, j\big\rangle^{(1)}&=&\!\!\int_w \frac{1}{2}\left( |w,\,_{i\, j}\rangle{+} |w,\,_{j\, i}\rangle\right) {-}\frac{\delta_{i j}}{D{-}1}|w, \,_{k\, k}\rangle,\quad\big| i, j\big\rangle^{(2)}\!=\!\!\int_w \left(w_i w_j{-}\frac{\delta_{i j}}{D{-}1}\right)|w, \,_{k\, k}\rangle,\nonumber\\
\big|{\tiny \begin{array}{c}i \\j \\k\end{array}}\big\rangle&=&\int_w w_{[i} |w,\,_{ j\, k]}\rangle\,,
\end{eqnarray}
where $\int_w$ stands for the projective integral over a $(D{-}2)$-sphere, i.e., $\int_w\equiv \int \langle w\, \d^{D{-}2}w\rangle$. Note that we have two possible spin-2 structures.\footnote{Other spin-2 structures will vanish when contracting with the polarization vectors. For example consider $\big| i, j\big\rangle^{(3)}=\int_w  w_{\{i} w_k |w,\,_{j\}\,k }\rangle$. Then 
\begin{equation}
E^a_{1}E^b_{2}\big\langle x,\,_{a\,b} \big| i, j\big\rangle^{(3)}=(E_{2}\cdot x)E_{1\{i}x_{j\}}\,
\end{equation}
which vanishes since $E_{2}\cdot x=0$.} As it turns out, this will be sufficient. The inner product with $|x, a\, b\rangle$ is given as
\begin{eqnarray}
\big\langle x,\,_{a\,b} \big|\cdot\big\rangle&=& \delta_{ab},\nonumber\\
\big\langle x,\,_{a\,b} \big| i, j\big\rangle^{(1)}&=&\frac{(\delta_{ai}\delta_{bj}{+}\delta_{aj}\delta_{bi})}{2}{-}\frac{\delta_{ij}\delta_{ab}}{D{-}1}\,,\quad \big\langle x,\,_{a\,b} \big| i, j\big\rangle^{(2)}=(x_i x_j{-}\frac{\delta_{ij}}{D{-}1})\delta_{ab}\,,\nonumber\\
\big\langle x,\,_{a\,b} \big|{\tiny \begin{array}{c}i \\j \\k\end{array}}\big\rangle&=&x_{[i}\delta_{aj}\delta_{bk]}\,.
\end{eqnarray}
Using these irreps, we can construct the following rotationally-invariant operator,
\begin{eqnarray}
\mathcal{O}&=&c_0\big| \cdot\big\rangle\big\langle \cdot\big|+
\sum_{\alpha,\beta=1}^{2} c_{\alpha \beta} \big| i, j\big\rangle^{(\alpha)}\big\langle i, j\big|^{(\beta)}
+c_3 \big|{\tiny \begin{array}{c}i \\j \\k\end{array}}\big\rangle\big\langle{\tiny \begin{array}{c}i \\j \\k\end{array}}\big|\,.
\end{eqnarray} 
Sandwiched between $|x,E_1,E_2\rangle=|x,\,_{a\,b}\rangle E^a_1E^b_2$ and $\langle y,E_3,E_4|$ we find
\begin{eqnarray}
&&\langle y,E_3,E_4|\mathcal{O}|x,E_1,E_2\rangle=(E_1{\cdot} E_2)(E_3{\cdot}  E_4)\left[c_0{-}\frac{c_{11}{+}c_{12}{+}c_{21}{+}c_{22}}{D-1}{+}(x{\cdot} y)^2 c_{22}\right]\nonumber\\
&&\qquad\qquad+(E_1{\cdot}  E_3)(E_2{\cdot}  E_4)\left[\frac{c_{11}}{2}{-}(x{\cdot}  y)c_{3}\right]{+}(E_2{\cdot}  E_3)(E_1{\cdot}  E_4)\left[\frac{c_{11}}{2}{+}(x{\cdot}y)c_{3}\right]\nonumber\\
&&\qquad\qquad+(E_3{\cdot} E_4)(E_1{\cdot}  y)(E_2{\cdot}  y)c_{12}{+}(E_1{\cdot}  E_2)(E_3{\cdot}  x)(E_4{\cdot} x)c_{21}\nonumber\\
&&\qquad\qquad+\Big[(E_2{\cdot} E_4)(E_1{\cdot}  y)(E_3{\cdot}  x){-}(E_2{\cdot}  E_3)(E_1{\cdot}  y)(E_4{\cdot}  x)\nonumber\\
&&\qquad\qquad\quad+(E_1{\cdot}  E_3)(E_2{\cdot}  y)(E_4{\cdot}  x){-}(E_1{\cdot}  E_4)(E_2{\cdot} y)(E_3{\cdot}  x)\Big]c_3\,.
\end{eqnarray}
The coefficients can now be fixed by matching to $\mathcal{F}^4$ in eq.~(\ref{F4Form}), leading to:
\begin{eqnarray}
c_0=\left(\frac{9}{D{-}1}{-}1\right),\;\;c_{11}=4,\;\;c_{12}=c_{21}=2,\;\; c_{22}=1,\;\;c_3=2\,.
\end{eqnarray}

Since the states are irreps, they are orthogonal, and unitarity demands $c_0\geq0$ which implies that $D\leq 10$. Thus tree-level Yang--Mills amplitude ``knows" about critical dimension of type-I string! Note that the matrix of couplings for the spin-2 state has rank one: 
\begin{equation}
   \left( \begin{array}{cc} c_{11} & c_{12} \\ c_{21} & c_{22} \end{array}\right) = \left(\begin{array}{cc}4 & 2 \\ 2 & 1\end{array}\right)=v v^T\quad {\rm with} \, \quad v=\left(\begin{array}{c}2  \\ 1 \end{array}\right)\,,
\end{equation} 
which tells us that we have exactly one linear combination of the two spin-2 states appearing here, which we can identify as $\big|\,{\tiny \ytableausetup{centertableaux}\ydiagram{2}}\,\big\rangle=2| i,j\rangle^{(1)}{+}| i,j\rangle^{(2)}$. Thus we conclude that $(\mathcal{F}^4)_{a,b,c,d}(x, y)=\langle y, \,_{c\,d}|\mathcal{O}|x,\,_{a\,b}\rangle$ with
\begin{equation}
\mathcal{O}=\left(\frac{9}{D{-}1}{-}1\right)\big| \cdot\big\rangle\big\langle \cdot\big|+\big|\,{\tiny \ytableausetup{centertableaux}\ydiagram{2}}\,\big\rangle\big\langle \,{\tiny \ytableausetup{centertableaux}\ydiagram{2}}\,\big|+\big|\,{\tiny \ytableausetup{centertableaux}\ydiagram{1,0+1,0+1}}\,\big\rangle\big\langle \,{\tiny \ytableausetup{centertableaux}\ydiagram{1,0+1,0+1}}\,\big| \,.
\end{equation}

Importantly, once it is established that $(\mathcal{F}^4)_{a,b,c,d}(x, y)$ is a positive operator, this can be extended to $(\mathcal{F}^4)_{a,b,c,d}(x, y) \cdot R_n^{\I}(x\cdot y)$ if $R_n^{\I}(x\cdot y)$ has a positive expansion on the Gegenbauer polynomial. To see this for each state in the Hilbert space $|\psi \rangle$, a state $|\psi; z, j\rangle$ defined through 
\begin{equation}
\langle x,\,_{ab}|\psi; z, j\rangle =\langle x,\,_{ab}|\psi \rangle\; G^{(D)}_j(x\cdot z)\,.
\end{equation} 
In other words, $|\psi; z, j\rangle$ is the result of tensoring  $|\psi \rangle$ with symmetric spin-$j$ irrep. Note that even if the $|\psi \rangle$'s are irreps, $|\psi; z, j\rangle$ in general will not be and thus are not orthogonal. Thus once we have
\begin{equation}
R_n^{\I}(x\cdot y)=\sum_j B_{n,j}^D\,G^{(D)}_j(x\cdot y),\quad B^D_{n,j}\geq0,
\end{equation}
we can construct $(\mathcal{F}^4)_{a,b,c,d}(x, y) \cdot R_n^{\I}(x\cdot y)$ as a positive operator:
\begin{eqnarray}
\mathbb{R}_n=\sum_j B_{n,j}^D\Bigg[&&\int_z\left( \frac{9}{D{-}1}{-}1\right)\big| \cdot; z, j\big\rangle\big\langle \cdot; z, j\big|\\
&&+\big|\,{\tiny \ytableausetup{centertableaux}\ydiagram{2}}; z, j\big\rangle\big\langle \,{\tiny \ytableausetup{centertableaux}\ydiagram{2}}; z, j\big|+\big|\,{\tiny \ytableausetup{centertableaux}\ydiagram{1,0+1,0+1}}\,; z, j\big\rangle\big\langle \,{\tiny \ytableausetup{centertableaux}\ydiagram{1,0+1,0+1}}\,; z, j\big|\Bigg]\,.\nonumber
\end{eqnarray}
We have already shown at $n=1$ we must have $D\leq 10$. Numerically, one observes that $B_{n,j}^D$ is indeed positive for $D\leq 10$. In the next section we will prove its positivity for $D\leq6$.

We note in passing that we can describe the usual Gegenbauer expansion for external scalars, in exactly this same positive-operator-in-Hilbert-space language. Here the states of the Hilbert space are simpler, just labeled by the unit vector on the sphere $|w \rangle$, and the residue $R_n(x \cdot y)$ defines an operator $\mathbb{R}_n$ via $\langle x| \mathbb{R}_n | y \rangle = R_n(x \cdot y)$. The only irreps of SO($D{-}1$) we can build from these are symmetric spin $j$ tensors, as $| \cdot \rangle = \int_w |w \rangle, |i_1\rangle = \int_w w_{i_1} |w \rangle, |i_1 i_2 \rangle = \int_w (w_{i_1} w_{i_2} - \frac{1}{D-1} \delta_{i_1 i_2}) |w \rangle$ etc. Since the states are completely symmetric it is natural to define $|z;j \rangle = z_{i_1} \cdots z_{i_j} |i_1 \cdots i_j \rangle$. These states define the Gegenbauer polynomials as $\langle w|z;j\rangle = G^{(D)}_j(w \cdot z)$. Any rotationally-invariant operator ${\cal O}$ can be written as $\sum_j c_j |i_1 \cdots i_j \rangle \langle i_1 \cdots i_j| = \sum_j c_j \int_z |z;j \rangle \langle z;j|$. In particular we can expand the operator $\mathbb{R}_n = \sum B^{D}_{n,j} \int_z |z;j\rangle \langle z;j|$. Since these states are irreps and orthogonal, for positivity we must have that $B^D_{n,j} \geq 0$, and so the function $R_n(x\cdot y)$ must have an expansion as a sum over Gegenbauer polynomials with all positive coefficients.

Let us finally turn to the closed superstring amplitudes. Having seem that the open string residues are identified with positive operators, it is easy to see that via the usual ``closed = open$^2$" connection \cite{Kawai:1985xq}, the closed string residues since $\mathcal{R}^4=(\mathcal{F}^4)^2$ and $R_n^{\rm II}=(R_n^{\rm I})^2$ are positive as well. Let us begin with 
\begin{equation}
(\mathcal{R}^4)_{a_1b_1a_2b_2a_3b_3a_4b_4}=(\mathcal{F}^4)_{a_1a_2a_3a_4}(\mathcal{F}^4)_{b_1b_2b_3b_4}\,,
\end{equation}
and introduce the Hilbert space $|x,\,_{a_1b_1a_2b_2}\rangle$. Now given any two $|\psi \rangle$, $|\chi\rangle$ in the YM Hilbert space, we can once again define $|\psi, \chi\rangle$ in the gravity Hilbert space, through
\begin{equation}
\langle x,\,_{a_1b_1a_2b_2}|\psi, \chi\rangle=\langle x,\,_{a_1b_1}|\psi\rangle \langle x,\,_{a_2b_2}|\chi\rangle\,.    
\end{equation}
Similarly, for any two operators $\mathcal{O}_{\rm YM}$, $\mathcal{O}'_{\rm YM}$ on the YM Hilbert space,
\begin{equation}
    \mathcal{O}_{\rm YM}=\sum_\psi c_\psi |\psi \rangle\langle \psi|,\quad \mathcal{O}'_{\rm YM}=\sum_\chi c_\chi |\chi \rangle\langle \chi|\,,
\end{equation}
we define an operator on the gravity Hilbert space $
    \mathcal{O}^{\rm gr}_{\mathcal{O}_{\rm YM}\mathcal{O}'_{\rm YM}}=\sum_{\chi,\psi} c_\psi c_\chi |\psi,\chi\rangle\langle \psi,\chi|$ with
    \begin{eqnarray}
\langle y,\,_{a_3b_3a_4b_4}| \mathcal{O}^{\rm gr}_{\mathcal{O}_{\rm YM}\mathcal{O}'_{\rm YM}}|x,\,_{a_1b_1a_2b_2}\rangle&=&\sum_{\psi,\chi}c_\psi c_\chi \langle y,\,_{a_3a_4}|\psi\rangle\langle \psi|x,\,_{a_1a_2}\rangle\langle y,\,_{b_3b_4}|\chi\rangle\langle \chi|x,\,_{b_1b_2}\rangle\nonumber\\
&=&\langle y,\,_{a_3a_4}|\mathcal{O}_{\rm YM}|x,\,_{a_1a_2}\rangle\langle y,\,_{b_3b_4}|\mathcal{O}'_{\rm YM}|x,\,_{b_1b_2}\rangle\,.
    \end{eqnarray}
Thus we have 
\begin{equation}
(\mathcal{R}^4)_{a_1b_1a_2b_2a_3b_3a_4b_4}=\langle y,\,_{a_3b_3a_4b_4}|\left(\sum_{\psi,\psi'}c_\psi c_{\psi'}|\psi,\psi'\rangle\langle \psi,\psi'| \right)|x,\,_{a_1b_1a_2b_2}\rangle\,,
\end{equation}
where $\big|\psi\big\rangle=(\big| \cdot\big\rangle, \; \big|\,{\tiny \ytableausetup{centertableaux}\ydiagram{2}}\,\big\rangle,\; \big|\,{\tiny \ytableausetup{centertableaux}\ydiagram{1,0+1,0+1}}\,\big\rangle)$, $c_{ \cdot}=\left(\frac{9}{D{-}1}{-}1\right)$ and all the remaining $c_\psi$ are 1. Thus we see that for $D\leq 10$, since all $c_\psi$ are non-negative, once again we have a positive operator. Note that for $D>10$, the coefficients for $(\big| \cdot,\;{\tiny \ytableausetup{centertableaux}\ydiagram{2}}\,\big\rangle, \big|\,\cdot,\;{\tiny \ytableausetup{centertableaux}\ydiagram{1,0+1,0+1}}\,\big\rangle)$ are negative. This however, does not imply negativity since the states are not orthogonal, and may have overlap. This is indeed the case, as seen from the fact that the 3-form  $\big|\,\cdot,\;{\tiny \ytableausetup{centertableaux}\ydiagram{1,0+1,0+1}}\,\big\rangle$ can also be generated from $\big|\,{\tiny \ytableausetup{centertableaux}\ydiagram{2}},\;{\tiny \ytableausetup{centertableaux}\ydiagram{1,0+1,0+1}}\,\big\rangle)$. Thus generically the ``critical dimension" for closed string amplitudes---at least as directly inferred from our unitarity check for four-point scattering---are higher than their open string counterparts.

\section{\label{sec:proof}Double-contour representation}

We now turn to deriving our central result, the new double-contour integral representation \eqref{eq:double contour formula} for the coefficients of the partial wave expansion of the residue polynomials defined above. In Appendix~\ref{app:single variable generating function}, we give an alternative derivation of the same formula.

To begin with, let us consider the type-I amplitude, for which 
\begin{equation}
A^\I(s,t) = \frac{\Gamma({-}s) \Gamma({-}t)}{\Gamma(1{-} s {-} t)} = {-}\frac{1}{s}
\int_0^1 \d z\, z^{{-}s} (1{-}z)^{{-}1{-}t}\,.
\end{equation}
We would like to extract the residues on the massive poles of this expression at $s=n$, $A^\I(s,t) \to \frac{1}{s-n} \times R^\I_n(t)$ for $n=1,2,\ldots$. Of course we can trivially find from the explicit form of the Gamma functions that $R^\I_n(t) = \frac{1}{n!}(1+t)(2+t) \cdots (n-1 + t)$, but we would like to represent this slightly more naturally directly from the integral representation. To whit, note that the the singularities in $s$ all come from the region of integration near $z \to 0$. This motivates Taylor expanding the integrand around $z=0$, which as is familiar gives us the expansion of the amplitude as a sum over poles in $s$. Putting $(1 - z)^{-1 - t} = \sum_m a_m(t) z^m$, we have $A^\I(s,t) = \sum_m \frac{a_m(t)}{s - m +1}$, so that the residue $R^\I_n(t) = a_{n-1}(t)$. We can in turn represent $a_{n-1}(t)$ as $n{-}1$ derivatives around $z=0$ of the function $(1 - z)^{-1-t}$, or what is the same, as a contour integral around $z=0$, as $R^\I_n(t)=\frac{1}{n} \oint_{z=0} \frac{\d z}{2\pi i\, z^n}\; (1{-}z)^{{-}1{-} t}$. For the purpose of the partial wave expansion, we would like to write this expression in terms of the $x= \cos\theta$, which is related to $t$ as usual $t = s(x -1)/2 = n(x -1)/2$ where we have used that on the resonance $s=n$. Finally, since the dependence on $x$ is exponential via $(1 - z)^{-nx/2}$, it is useful to introduce $(1 - z) = e^{-u}$ to manifest this dependence. The contour integral around $z=0$ translates into one around $u = 0$, and we have for the residue polynomial $R_n^\I(x)$: 
\begin{equation}
    R^\I_n(x) =\frac{1}{n} \oint_{u=0} \frac{\d u}{2\pi i} \frac{e^{\frac{n}{2} u} e^{\frac{n}{2} u x}}{(e^u - 1)^n}\,.
\end{equation}
Note that this expression transforms simply under $u \to -u$, $x \to -x$, and we discover that $R^\I_n(-x) = (-1)^{n+1} R^\I_n(x)$, as we have already observed for the residue polynomial. 

We would now like to expand $R_n^\I(x)$ in the basis of Gegenbauer polynomials $G_j^{(D)}(x)$, which are orthogonal with respect to the usual measure $\d \theta\, (\sin \theta)^{(D-3)} = \d x\, (1 - x^2)^{\frac{D-4}{2}}$ (compared to the classic conventions \cite{watson1995treatise}, we have $G_j^{(D)}(x) = C_j^{(\alpha)}(x)$ with $\alpha = \tfrac{D-3}{2}$). For convenience given its ubiquitous appearance in our expressions, we will define $\delta = \frac{D - 4}{2}$. We will also use the Rodrigues formula for $G_j^{(D)}(x)$: 
\begin{equation}
    G_j^{(D)}(x) = (-1)^j\alpha_{j,D} (1 - x^2)^{-\delta}\, \partial_x^j (1 - x^2)^{\delta + j}.
\end{equation}
where $\alpha_{j,D}=(2\delta{+}1)_j/(2^j\,j!\, (\delta{+}1)_j)$. 
Now by orthogonality, any function $F(x)$ can be expanded in Gegenbauers as $F(x) =  \sum_{j=0}^\infty F_j G_j^{(D)}(x)$, where
\be 
F_j =n_{j,D} \int_{-1}^1 \d x\, (1 - x^2)^{\delta}\, G_j^{(D)}(x) F(x)\ , \qquad 
n_{j,D}=j!\, \frac{2^{{2}\delta}(j{+}\delta{+}\frac{1}{2})[\Gamma(\delta{+}\frac{1}{2})]^2}{\pi \Gamma(j{+}2\delta{+}1)}\ .
\ee
We will suppress these positive factors for now for simplicity, with the understanding that an additional factor of $\alpha_{j,D}n_{j,D} $ is to be restored to the final answer.   

Using the Rodrigues formula, and assuming that $F(x)$ is regular at $x=\pm 1$, we can integrate by parts $j$ times to arrive at $F_j = \int_{-1}^1 \d x\, (1 - x^2)^{\delta + j} \partial^j_x F(x)$. 
We can apply this form to extract the coefficients in the Gegenbauer expansion of the residue polynomial $R_n^\I(x)$. We have
\begin{equation}
    R_n^\I(x) = \sum_j B^D_{n,j} G_j^{(D)}(x),
\end{equation}
where 
\begin{eqnarray}
    B^D_{n,j} &=& \int_{-1}^1 (1 - x^2)^{\delta + j} \,\partial_x^j \oint_{u=0} \frac{\d u}{2\pi i} \frac{e^{\frac{n}{2} u} e^{\frac{n}{2} u x}}{(e^u - 1)^n} \nonumber \\ 
    &=& \frac{1}{n}\left(\frac{n}{2}\right)^j\oint_{u=0} \frac{\d u}{2\pi i} \frac{e^{\frac{n}{2} u} u^j}{(e^u - 1)^n} \int_{-1}^1 \d x\, (1 - x^2)^{\delta + j} e^{\frac{n}{2} u x}\,.
    \end{eqnarray}
Now we can easily perform the $\int_{-1}^{1} \d x$ integral above, using
\begin{eqnarray}
    \int_{-1}^1 \d x\, (1 - x^2)^J e^{a x} &=& (1 - \partial_a^2)^J \int_{-1}^1 \d x\, e^{a x} \nonumber = (1 - \partial_a^2)^J \left(\frac{e^{a} - e^{-a}}{a} \right) \nonumber \\ & = & H_J(a) + H_J(-a),
    \end{eqnarray}
where we used the notation $J = j+\delta$ and
\begin{equation}
    H_J(a) \equiv (1 - \partial_a^2)^J \left(\frac{e^{a}}{a} \right)\,.
\end{equation}
Note that here we need to specialize to even $D$ so that $J$ is an integer.
Thus we find
\begin{equation}
    B^D_{n,j} =\tfrac{1}{n}\left(\tfrac{n}{2}\right)^j \oint_{u=0} \frac{\d u}{2\pi i} \frac{e^{\frac{n}{2} u} u^j}{(e^u - 1)^n} \left[H_{\delta{+}j}\left(\tfrac{n}{2}u\right) + H_{\delta{+}j}\left({-}\tfrac{n}{2}u\right) \right].
\end{equation}
Now, the same parity symmetry $u \to - u$ we used above in establishing the parity of the residue polynomial $R_n^\I(x)$, tells us here that the integral with the first term $H_{\delta{+}j}(\frac{n}{2} u)$ is $(-1)^{n+j+1}$ times that of the second term $H_{\delta{+}j}(-\frac{n}{2} u)$. We thus learn that $B^D_{n,j}$ vanishes when $n{+}j$ is even, while while $n{+}j$ is odd we have 
\begin{equation}
    B^D_{n,j} =\left(\tfrac{n}{2}\right)^{j{-}1}\oint_{u=0} \frac{\d u}{2\pi i} \frac{e^{\frac{n}{2} u} u^j}{(e^u - 1)^n} H_{\delta{+}j}\left(\tfrac{n}{2}u\right)\,.
\end{equation}

Up to this point, all of the steps in our derivation have been completely straightforward, obvious and natural. But now something more slightly magical happens. The object $H_J(a) = (1{-}\partial_a^2)^J (e^{a}/a)$ we have introduced does not appear especially simple, so that we still do not have a good way of integrating against this factor. But, in fact, $H_J(a)$ is essentially a $J$-th total derivative of a simple object, that will allow us to proceed with the integration in a simple way. In fact we have a remarkable identity, 
\begin{equation}\label{ID1}
(1{-}\partial_a^2)^J \left(\frac{e^{a}}{a} \right) = J!\, e^{{-}a}\, \partial_a^J \left(\frac{e^{2 a}}{a^{J{+}1}} \right)\,.
\end{equation}

With this identity, we immediately arrive at our double contour-integral representation. Using this representation for $H_J(\frac{n}{2} u)$ and integrating by parts $J$ times we have 
\begin{equation}
B^D_{n,j} = \left(\tfrac{n}{2}\right)^{{-}2\delta{-}2{-}j}(-1)^{j{+}\delta}(j{+}\delta)!\oint_{u=0} \frac{\d u}{2\pi i} \frac{e^{n u}}{u^{j{+}\delta{+}1}} \partial_u^{j {+} \delta} \left( \frac{u^j}{(e^u {-} 1)^n} \right)
\end{equation}
and we can finally write the $(j{+}\delta)$-th derivative above as a contour integral, using $\partial_u^J F(u) = (-1)^J J! \oint_{v = 0} \frac{\d v}{2\pi i\, v^{J{+}1}} F(u{-}v)$, to obtain our double-contour integral
\begin{eqnarray} 
B^D_{n,j} &=&\left(\tfrac{n}{2}\right)^{{-}2\delta{-}2{-}j} [(j{+}\delta)!]^2 \oint_{u=0} \frac{\d u}{2\pi i} \frac{e^{n u}}{u^{j{+}\delta{+}1}} \oint_{v=0} \frac{\d v}{2\pi i} \frac{(u{-}v)^j}{v^{j{+}\delta{+}1} (e^{(u{-}v)}{-} 1)^n} \nonumber \\ &=&\left(\tfrac{n}{2}\right)^{{-}2\delta{-}2{-}j} [(j{+}\delta)!]^2 \oint_{u=0} \frac{\d u}{2\pi i} \oint_{v=0} \frac{\d v}{2\pi i} \frac{(v-u)^j}{(u v)^{j {+} \delta {+} 1} (e^v {-} e^u)^n}\,.
\end{eqnarray}
Finally, restoring the extra positive factor of $\alpha_{j,D} n_{j,D}$, we have
\begin{equation}\label{eq:B to beta}
B^D_{n,j}=c_{n,j}^D  \beta^{D}_{n,j}, \qquad c_{n,j}^D=\frac{2^{D-2}\Gamma(\frac{D-3}{2})(j+\frac{D-3}{2})(j+\frac{D-4}{2})!}{\sqrt{\pi}\,n^{j+D-2}},
\end{equation}
where $\beta^{D}_{n,j}$ is the double-contour integral displayed on the right-hand side of \eqref{eq:double contour formula} with $a=0$.

All that remains is to naturally motivate and derive the fundamental identity in eq. (\ref{ID1}). The motivation is simple: in order to act with the operator $(1 - \partial_a^2)^J$, it is natural to work with Fourier/Laplace representations that diagonalize $\partial_a$. The Laplace transform is obviously suggested, since $\frac{1}{a} = \int_0^\infty \d q\, e^{-q a}$. Differentiating $J$ times we also have that $ J!/a^{J+1} = \int_0^\infty \d q\, q^J e^{-q a}$. Putting $q = r{-}p$ for any real $p$, we find the Laplace representation
\begin{equation}
    \frac{J!\, e^{{-}p a}}{a^{J{+}1}} = \int_p^\infty \d r\, (r{-}p)^J e^{{-}r a}.
\end{equation}
Now, let us begin with this representation for $e^{-a}/a = \int_1^\infty \d t\, e^{-t a}$. In this form we can trivially apply $(1 - \partial_a^2)^J$ to get
\begin{equation}
    (1 - \partial_a^2)^J \left(\frac{e^{-a}}{a} \right) = \int_1^\infty \d t \ (1 - t^2)^J e^{-t a}.
\end{equation}
It is now obvious that we should recognize this object at a $J$-th total derivative. The reason is that $(1 - t^2)^J = (1-t)^J (1+t)^J$ factorizes, so it is natural to put $1 {+} t = r$, and discover our identity:
\begin{eqnarray} 
(1 - \partial_a^2)^J \left(\frac{e^{-a}}{a}\right) &=& e^a \int_2^\infty \d r\, (-r)^J (r - 2)^J e^{-r a} \nonumber \\ &=& e^a\, \partial_a^J \int_2^\infty \d r (r - 2)^J e^{-r a} \nonumber \\
&=& J!\, e^a\, \partial_a^J \left( \frac{e^{-2 a}}{a^{J+1}} \right)\,.
\end{eqnarray}
Changing $a\rightarrow {-}a$ yields eq. (\ref{ID1}).
\medskip

We can also find a double-contour representation of the residues for scattering the $m^2 = -1$ tachyon states in the bosonic string, beginning with the Veneziano amplitude $A^{\open,\bosonic}(s,t) = \Gamma({-}1{-}s) \Gamma({-}1{-} t)/\Gamma({-}2{-}s{-}t)$. Following the same steps as in the above derivation, we find   
\begin{equation}\label{eq:B bosonic}
    B^{\open, \bosonic}_{n,j} ={c'}^D_{n,j} \oint_{u=0} \frac{\d u}{2\pi i} \oint_{v=0} \frac{\d v}{2\pi i} \frac{(v{-}u)^j e^{{-}(u{+}v)}}{(u v)^{j{+}\delta{+}1} (e^v{-} e^u)^{n{+}2} }\,,
\end{equation}
where ${c'}^D_{n,j}=2^{D-2}\Gamma(\frac{D-3}{2})(j+\frac{D-3}{2})(j+\frac{D-4}{2})!/(\sqrt{\pi}(n+4)^{j+D-3})$. We will refer to the part stripped off from this constant as $\beta^{\open,\bosonic}_{n,j}$, just as in \eqref{eq:double contour formula} with $a=1$. See Appendix~\ref{app:triple} for a step-by-step derivation.

Similar formula for partial-wave expansion coefficients can be obtained for closed strings. We parametrize their amplitudes with
\be\label{eq:A-closed}
A^{\closed}_{a,b}(s,t) := \tfrac{1}{\pi} s^{a + b - 2} \int_{\mathbb{C}} \d^2 z\, \frac{|z|^{-2s} |1-z|^{-2t}}{z^{2a} (1-z)^{1+a} \bar{z}^{2b} (1-\bar{z})^{1+b}},
\ee
for $a,b \in \mathbb{Z}$ and $\d^2 z = \frac{1}{2i} \d z \wedge \d\bar{z}$.
Explicitly, we have
\be\label{eq:closed-cases}
A^{\closed}_{a,b}(s,t) = \begin{dcases}
	\quad\frac{\Gamma(-s)\Gamma(-t)\Gamma(-u)}{\Gamma(1{+}s)\Gamma(1{+}t)\Gamma(1{+}u)} &\quad\text{for}\quad a=b=0\qquad \text{(type-II)},\\
	\frac{\Gamma({-}s{-}1) \Gamma({-}t{-}1) \Gamma({-}u{+}3)}{\Gamma(1 {+}s) \Gamma(1{+}t) \Gamma(1{+}u)} &\quad\text{for}\quad a=0, b=1 \quad \text{(heterotic)},\\
	-\frac{\Gamma(-1{-}s)\Gamma(-1{-}t)\Gamma(-1{-}u)}{\Gamma(2{+}s)\Gamma(2{+}t)\Gamma(2{+}u)} &\quad\text{for}\quad a=b=1\qquad \text{(bosonic)},
	\end{dcases}
\ee
where $u = -s-t + 4m^2$. We have $m^2 = -\min(a,b)$, i.e., scattering of gravitons in the superstring, gluons in the heterotic, and tachyons in the bosonic one.\footnote{Once again there will be an additional overall $\delta^{2\times 8}(Q)$ for the full type-II amplitude, and $\delta^{8}(Q)$ for the heterotic amplitude.} In this notation, the spins range in $j \leq 2(n{-}1{+}a{+}b)$. Note that using Kawai--Lewellen--Tye relations \cite{Kawai:1985xq} we can express it in terms of the open-string amplitudes
\be
A_{a,b}(s,t) = \frac{1}{\pi} \frac{\sin (\pi s)\, \sin (\pi t)}{\sin (\pi(s{+}t))} A_{a}(s,t) A_b(s,t),
\ee
which implies that
\be\label{eq:closed-residue}
\Res_{s=n} A_{a,b}(s,t) = \left( \Res_{s=n} A_{a}(s,t) \right) \left( \Res_{s=n} A_{b}(s,t) \right).
\ee
(An alternative, more direct derivation is given in eq.~\eqref{eq:C1}.) Therefore positivity of the Gegenbauer expansion of closed-string amplitudes follows from that of the open case. Nevertheless, it is still interesting to find a closed-string counterpart, which we denote as $\beta_{n,j}^D$,  of the double-residue formula. We specialize to $D \in 2\mathbb{Z}$.

Following a similar set of manipulations to those in the previous subsection, one finds in the cases $a=b$ that $B_{n,j}^{\closed,\II/\bosonic} = 0$ for odd $j$ (for any $n$ and $D$), while for even $j$ we have
\begin{align}
B_{n,j}^{\closed,\II/\bosonic} = 2\, n^{2(a-1)}\, &[(j{+}\tfrac{D-4}{2})!]^2 \oint_{u=0} \frac{\d u}{2\pi i} \oint_{\tilde{u}=0} \frac{\d \tilde{u}}{2\pi i} \oint_{v=0} \frac{\d v}{2\pi i} \nn\\
& \qquad\qquad\times \frac{e^{v n + (4v-u-\tilde{u})a } (u + \tilde{u})^j}{[(1-e^u)(1-e^{\tilde{u}})]^{n+2a} [v(v{-}u{-}\tilde{u})]^{j+\frac{D-2}{2}}},\label{eq:closed-super}
\end{align}
where the residue is taken around the origin in all the variables. In the heterotic string case, $a=0$, $b=1$, there is no distinction between even/odd $j$ and we have
\begin{multline}
B_{n,j}^{\closed,\het} = \!\tfrac{1}{n}\, [(j{+}\tfrac{D-4}{2})!]^2 \\
\times \oint_{u=0} \frac{\d u}{2\pi i} \oint_{\tilde{u}=0} \frac{\d \tilde{u}}{2\pi i} \oint_{v=0} \frac{\d v}{2\pi i} \frac{((-1)^j -  e^{4\tilde{u}})\, e^{- \tilde{u} + vn} (u + \tilde{u})^j}{(1{-}e^u)^{n} (1{-}e^{\tilde{u}})^{n+2} [v(v{-}u{-}\tilde{u})]^{j+\frac{D-2}{2}}}\,.\label{eq:closed-heterotic}
\end{multline}
Proofs of these formulae are given in Appendix~\ref{app:triple}, where one can also find alternative derivations and expressions in terms of quadruple-contour integrals.

\section{\label{sec:applications}Applications}
In this section, we show that our double contour formula \eqref{eq:double contour formula} is well-suited to study questions of unitarity. From now on, we use $\beta_{n,j}^{D}$ to denote the double-contour representation stripped off from the manifestly-positive constant according to \eqref{eq:B to beta}, and similarly for the bosonic case.

\subsection{\label{sec:manifest unitarity}Manifest unitarity in \texorpdfstring{$D \le 6$}{D <= 6} dimensions}
The double contour formula makes unitarity manifest in $D \le 6$ spacetime dimensions for the case of the superstring. To see this, we simply have to change variables to 
\be 
u=\log(1-x)\ , \qquad v=\log(1-y)\ ,
\ee
in terms of which the integral becomes
\begin{multline} 
\beta_{n,j}^D =\oint_{x=0} \frac{\mathrm{d}x}{2\pi i} \oint_{y=0} \frac{\mathrm{d}y}{2\pi i} \ \frac{1}{(1-x)(1-y)(\log(1-x)\log(1-y))^{\frac{D{-}2}{2}}} \frac{1}{(x-y)^{n-j}}\\
\times \left(\frac{\frac{1}{\log(1-x)}-\frac{1}{\log(1-y)}}{x-y}\right)^j\ . \label{eq:double contour integral log form}
\end{multline}
Each of the three factors that enter the integrand is a function with only positive Taylor coefficients when expanded first around $y=0$ and then around $x=0$. We call such a function (or rather power series) a \emph{positive} function. Consequently, the residue integral that just picks out one of the coefficients will be positive.

Indeed, the factors,
\be 
\frac{1}{(1-x)(-\log(1-x))^{\frac{D{-}2}{2}}}
\ee
are of the same form as the function $h_\alpha$ for $\alpha=\frac{2{-}D}{2}$ discussed in the Appendix~\ref{app:positive function}. They are positive for $\alpha \ge -2$, i.e.,\ $D \le 6$.
The expansion of the function 
\be 
(x-y)^{-k}=\sum_{m=0}^\infty \binom{k+m-1}{m} y^{m}x^{-k-m}
\ee
is obviously also positive (since $k=n-j \ge 0$). Finally, the positivity of the last factor follows from the positivity of
\be 
f(z)=\frac{1}{\log(1-x)}+\frac{1}{x}=\sum_{m=0}^\infty c_m x^m
\ee
with $c_m \ge 0$. We demonstrate that also in Appendix~\ref{app:positive function} that this function is positive. Writing
\begin{align} 
\frac{\frac{1}{\log(1-x)}-\frac{1}{\log(1-y)}}{x-y}&=\frac{f(x)-\frac{1}{x}-f(y)+\frac{1}{y}}{x-y} \nonumber\\
&=\frac{1}{xy}+\sum_{m=0}^\infty c_m \frac{x^m-y^m}{x-y} \nonumber\\
&=\frac{1}{xy}+\sum_{m=0}^\infty c_m \sum_{k=0}^{m-1} x^{m-1-k} y^k 
\end{align}
shows that all coefficients are positive in the expansion. Thus in \eqref{eq:double contour integral log form} the integrand is a positive function and so in particular also the residue is positive. Thus our formula makes unitarity manifest for $D\le 6$ spacetime dimensions.

The same arguments also apply for the bosonic string. The double contour formula in its logarithmic form has an extra $\frac{1}{(1-x)(1-y)}$ compared to \eqref{eq:double contour integral log form} (as well as $n \to n+2$ replaced). Our formula makes then unitarity manifest for $D \le 10$ spacetime dimensions, because the function
\be 
\frac{1}{(1-x)^2 (-\log(1-x))^{\frac{D-2}{2}}}
\ee
is the square of the positive function $h_\alpha$ discussed in Appendix~\ref{app:positive function} with $\alpha=\frac{2-D}{4}$. This function is positive for $\alpha \ge -2$, i.e.,\ $D \le 10$.

\subsection{\label{sec:unitarity in D=10}Unitarity in \texorpdfstring{$D=10$}{D=10} for low spin}
While our formula does not make unitarity manifest in ten spacetime dimensions, one can still use it to demonstrate unitarity for low values of spin. We will give the argument here for $j=0$. Higher values of $j$ and the corresponding analogue for the bosonic string can also be checked on a case-by-case basis, but the computation gets more and more involved.

For $j=0$, starting with eq. (\ref{eq:double contour integral log form}) one can simply compute the $y$-residue and get
\be
\beta_{n,0}^{10} = \oint_{x=0} \frac{\mathrm{d}x}{2\pi i}\frac{n \left((n{+}2)(n{+}1){-}3(n{+}1)x{+}x^2\right)}{6 (1-x) x^{n+3} \log(1-x)^4}\ .
\ee
One can again reduce the statement $\beta^{10}_{n,0} \ge 0$ to the positivity of a particular generating function as follows. We can trade the $n$-dependence for derivatives and integrate by parts
\begin{align} 
\frac{\beta_{n,0}^{10}}{n(n{+}1)}&=\oint_{x=0} \frac{\mathrm{d}x}{2\pi i}\frac{1}{(n{+}1)} \partial_x \left(\frac{1}{18 \log(1-x)^3} \right)\left(\partial_x^2+3 \partial_x+1 \right)\left(x^{-n-1}\right) \nonumber\\
&=\oint_{x=0} \frac{\mathrm{d}x}{2\pi i} \frac{1}{(n{+}1)}\left(-\partial_x^2+3\partial_x-1 \right)\left(\frac{1}{18 \log(1-x)^3} \right) \partial_x \left(x^{-n-1}\right)\nonumber\\
&=\oint_{x=0} \frac{\mathrm{d}x}{2\pi i\, x^{n+2}} \left(\partial_x^2-3\partial_x+1 \right)\left(\frac{1}{18 \log(1-x)^3} \right) \nonumber\\
&=\oint_{x=0} \frac{\mathrm{d}x}{2\pi i\, x^{n+2}} \ \left(\frac{(1-x)^2 \log(1-x)^2+(9 x-6) \log(1-x)+12}{18 (1-x)^2 \log(1-x)^5}\right) \ .
\end{align}
Thus, it suffices to show that all Taylor coefficients starting from $\mathcal{O}(z)$ are positive for the following function
\begin{align} 
f(z)&=\frac{(1-z)^2 \log(1-z)^2+(9 z-6) \log(1-z)+12}{18 (1-z)^2 \log(1-z)^5} \\
&=-\frac{2}{3 z^5}+\frac{7}{18 z^3}-\frac{1}{36 z}+\frac{1}{2160}+\frac{z}{9072}+\frac{z^2}{20160}+\frac{z^3}{362880}+\frac{3077 z^5}{79833600}\nonumber\\
&\qquad\qquad+\frac{3125 z^6}{28740096}+\frac{13180873z^7}{65383718400}+\mathcal{O}(z^8)
\end{align}
Since unitarity is no longer manifest in our formula in $D=10$ dimensions, the singular coefficients in the expansion never enter the computation of $\beta_{n,0}^{10}$ and can be negative. We also remark that the $z^4$ term is absent, which indicates that $D=10$ is the critical dimension. We prove the positivity of the non-singular terms of this function in Appendix~\ref{app:positive function}.
\subsection{\label{sec:asymptotics}Asymptotics}
Finally, our double-contour formula is very useful to study asymptotics. 

\subsubsection{Fixed spin}
We again start from the form \eqref{eq:double contour integral log form}. In the large $n$ limit for fixed $j$, we can explicitly compute the residue in $y$. When computing the residue, we get terms with different powers of $n$. The greatest contribution is achieved by picking the highest possible power from the term $(x-y)^{-n+j}$, because these coefficients grow very rapidly with $n$. This means that we pick up the leading coefficients of the other factors. We hence get
\be 
\beta_{n,j}^{D} \sim \frac{n^{j+\frac{D{-}4}{2}}}{(j+\frac{D{-}4}{2})!} \oint_{x=0} \frac{\mathrm{d}x}{2\pi i} \ \frac{x^{-n-j-\frac{D{-}4}{2}}}{(1-x)(-\log (1-x))^{\frac{D{-}2}{2}}}\ ,
\ee
where we further approximated the binomial coefficient appearing in the expansion of $(x-y)^{-n+j}$ with its asymptotic value. We hence see in particular that the $j$-dependence of the asymptotic behaviour of the coefficients is very simple.

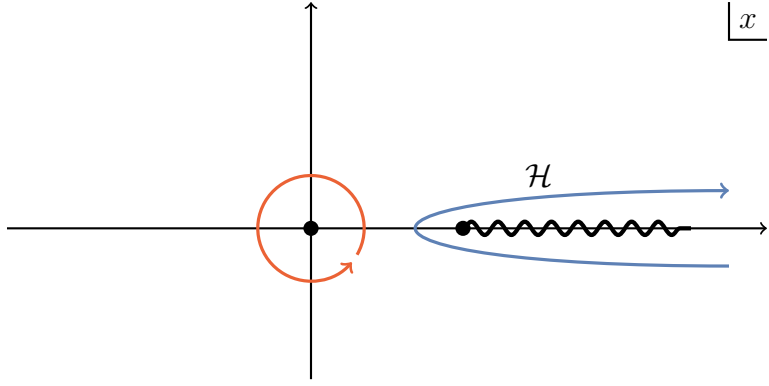
\begin{figure}
\begin{center}
\begin{tikzpicture}
\draw[thick,->] (-4,0) -- (6,0); 
\draw[thick,->] (0,-2) -- (0,3); 
\draw[thick] (5.5,3) -- (5.5,2.5) -- (6,2.5);
\node at (5.75,2.75) {$x$};
\fill (0,0) circle (.1);
\fill (2,0) circle (.1);
\draw[very thick,MathematicaRed,->] (.606,-.35) arc (-30:320:.7); 
\draw[very thick,MathematicaBlue,->] (5.5,-.5) .. controls (0,-.5) and (0,.5) .. (5.5,.5);
\node at (3,.7) {$\mathcal{H}$};
\draw[ultra thick,decorate,decoration={snake}] (2,0) -- (5,0);
\end{tikzpicture}
\end{center}
\caption{\label{fig:contour deformation}Contour deformation to the Hankel contour $\mathcal{H}$.} 
\end{figure}

It remains to determine the asymptotic value of the remaining integral. This can be done in a very classical way as follows, see e.g.\ \cite[Sec.~VI]{Flajolet}. We first deform the contour that runs around 0 to the Hankel contour that runs from below once around the branch cut as depicted in Figure~\ref{fig:contour deformation}. Since the integrand decays sufficiently fast at infinity, the arcs at infinity are not contributing to the integral. 
We next perform the change of variables $x=1+\frac{t}{n}$. The $t$ variables runs then from below the positive real axis, encircles the origin and then runs back above the real axis. We get
\begin{align} 
\beta_{n,j}^{D} &\sim -\frac{n^{j+\frac{D{-}4}{2}}}{(j+\frac{D{-}4}{2})!} \oint_{\mathcal{H}} \frac{\mathrm{d}t}{2\pi i} \ \frac{ \left(1+\frac{t}{n}\right)^{-n-j-\frac{D{-}4}{2}}}{t \left(-\log \left(-\frac{t}{n}\right)\right))^{\frac{D{-}2}{2}}}\\
&= -\frac{n^{j+\frac{D{-}4}{2}}}{(j+\frac{D{-}4}{2})! \log(n)^{\frac{D{-}2}{2}}}\oint_{\mathcal{H}} \frac{\mathrm{d}t}{2\pi i} \ \frac{\left(1+\frac{t}{n}\right)^{-n-j-\frac{D{-}4}{2}}}{t\left(1-\frac{\log (-t)}{\log(n)}\right)^{\frac{D{-}2}{2}}}\ .
\end{align}
The remaining integrand converges uniformly for any bounded domain of the $t$-plane and hence we can replace it with its value at large $n$, which gives
\be 
\beta_{n,j}^{D} \sim  -\frac{n^{j+\frac{D{-}4}{2}}}{(j+\frac{D{-}4}{2})! \log(n)^{\frac{D{-}2}{2}}} \oint_{\mathcal{H}} \frac{\mathrm{d}t}{2\pi i} \frac{\mathrm{e}^{-t}}{t}\ .
\ee
The contour integral finally simply picks up minus the residue, since we run clockwise around the pole. Thus, we get the simple, but interesting asymptotics
\begin{tcolorbox}
\be 
\text{fixed $j$,\quad large $n$:}\qquad \beta_{n,j}^{D} \sim  \frac{n^{j+\frac{D{-}4}{2}}}{(j+\frac{D{-}4}{2})! \log(n)^{\frac{D{-}2}{2}}} \ . \label{eq:fixed spin asymptotic}
\ee
\end{tcolorbox}\noindent
For the actual residue $B_j^n$, this means
\be 
B_{n,j}^{D} \sim \frac{2^{D{-}2}(j+\frac{D{-}3}{2})\Gamma\left(\frac{D{-}3}{2}\right)}{\sqrt{\pi}\, n^{\frac{D}{2}}\, \log(n)^{\frac{D{-}2}{2}}}\ .
\ee

In particular, the asymptotic values are positive.
This estimate can be extended to a whole asymptotic series in $\log(n)^{-1}$ by expanding the integrand in $\log(n)^{-1}$, which gives
\begin{align} 
\beta_{n,j}^{D} &\sim -\frac{n^{j+\frac{D{-}4}{2}}}{(j+\frac{D{-}4}{2})!} \sum_{m=0}^\infty \binom{\frac{D{-}4}{2}+m}{m} \frac{1}{\log(n)^{m+\frac{D{-}2}{2}}}\oint_{\mathcal{H}} \frac{\mathrm{d}t}{2\pi i} \frac{\mathrm{e}^{-t}}{t} \log(-t)^m \\
&\sim \frac{n^{j+\frac{D{-}4}{2}}}{(j+\frac{D{-}4}{2})!} \sum_{m=0}^\infty \binom{\frac{D{-}4}{2}+m}{m} \frac{(-1)^m}{\log(n)^{m+\frac{D{-}2}{2}}}\left.\frac{\mathrm{d}^m}{\mathrm{d}\alpha^m} \right|_{\alpha=1}\oint_{\mathcal{H}} \frac{\mathrm{d}t}{2\pi i} (-t)^{-\alpha}\mathrm{e}^{-t}\\
&\sim \frac{n^{j+\frac{D{-}4}{2}}}{(j+\frac{D{-}4}{2})!} \sum_{m=0}^\infty \binom{\frac{D{-}4}{2}+m}{m} \frac{(-1)^m}{\log(n)^{m+\frac{D{-}2}{2}}}\left.\frac{\mathrm{d}^m}{\mathrm{d}\alpha^m} \right|_{\alpha=1} \frac{1}{\Gamma(\alpha)} \ ,
\end{align}
where in the last step we made use of Hankel's formula for the Gamma function. 

By comparing the $m=0$ and the $m=1$ term, we can estimate that our asymptotic value becomes good in the regime $n \gg \mathrm{e}^{\frac{\gamma(D{-}2)}{2}}$, where $\gamma \approx 0.57$ is the Euler--Mascheroni constant. Hence we expect the couplings $B_j^n$ to be positive for $n \gtrsim \mathrm{e}^{\frac{\gamma(D{-}2)}{2}}$ for all dimensions. This only leads finitely many coefficients that can potentially be negative and can be checked by hand. However, we were unable to turn this approximate inequality into a rigorous bound, which prevents us from mathematically proving positivity of $B_{n,j}^{D}$ also in $D=10$ spacetime dimensions.

Finally, we note that the corresponding formula for the bosonic string is
\begin{align} 
\text{fixed $j$,\quad large $n$:}\qquad \beta_{n,j}^{\open,\bosonic}  &\sim  \frac{n^{j+\frac{D{-}2}{2}}}{(j+\frac{D{-}4}{2})! \log(n)^{\frac{D{-}2}{2}}} \ , \\
B_{n,j}^{\open,\bosonic} &\sim\frac{2^{D{-}2} (j+\frac{D{-}3}{2})  \Gamma \left(\frac{D{-}3}{2}\right) }{\sqrt{\pi }n^{\frac{D{-}4}{2}}\log(n)^{\frac{D{-}2}{2}}}\ .
\end{align}

\subsubsection{\label{sec:Regge}Regge trajectories}
The second asymptotics we can look at is to keep $\Delta=n-j$ fixed and consider large $n$. This limit corresponds to moving in one fixed Regge trajectory. In this case, it is convenient to start from the form
\be 
\beta_{n,n-\Delta}^{D} =-\frac{1}{2}\oint_{u=0} \frac{\mathrm{d}u}{2\pi i}\oint_{v=u} \frac{\mathrm{d}v}{2\pi i}\ \frac{1}{(uv)^{\frac{D{-}2}{2}}} \left(\frac{u^{-1}-v^{-1}}{\mathrm{e}^v-\mathrm{e}^u} \right)^n \left(u^{-1}-v^{-1} \right)^{-\Delta}\ . \label{eq:Regge limit contour integral 1} 
\ee
of the double contour formula. Notice that the inner integral runs around $v=u$ and not $v=0$ as before. To obtain this equivalent form, one the usual relation of contours with a singularity at $u=v$
\be
\oint_{v=0} \oint_{u=0} - \oint_{u=0} \oint_{v=0} = \oint_{u=0} \oint_{v=u}\ .
\ee
The second contour on the left-hand side is the original contour, but both terms give an equal contribution due to the symmetry of the integrand (for $n+j$ odd). Hence the contour $\oint_{u=0} \oint_{v=u}$ leads to $-2$ times the original formula. This is also explained in Appendix~\ref{app:triple}, where it is also explained that in this form the formula is actually also valid for odd $D$.

Since we want to get the dominating piece in the large $n$ limit, we should pick up as many powers of $n$ from the second factor as possible when doing the integral over $v$. This means that we can expand this term in $v$ around $u$. It turns out that one has to go to the second order and hence gets
\be 
\frac{u^{-1}-v^{-1}}{\mathrm{e}^v-\mathrm{e}^u} =\frac{\mathrm{e}^{-u}}{u^2} \left(1-\frac{(u+2) (v-u)}{2
   u}+\frac{\left(u^2+6 u+12\right) (v-u)^2}{12 u^2}\right) +\mathcal{O}\left((v-u)^3\right)\ .
\ee
In order for the second term in the expansion to make an appreciable contribution to the integral in the large $n$ limit, the term of order $\mathcal{O}(v-u)$ has to become small. This means that the $u$-integral generically has a large degree of cancellation built in. Note however that we can choose the radius of the $u$-integral to be $2$, so that we are integrating through the critical point $u=-2$, which becomes a saddle point in the large $n$ limit. This in turn implies that we can replace $u$ with $-2$ in most places that are not singular when $u \to v$, $u \to -2$ or $u \to 0$. We can also replace $v$ by $u$ in the first factor of \eqref{eq:Regge limit contour integral 1}. Hence we get the large-$n$ approximation
\be 
\beta_{n,n-\Delta}^{D} \sim -\frac{1}{2}\oint_{u=0} \frac{\mathrm{d}u}{2\pi i}\oint_{v=u} \frac{\mathrm{d}v}{2\pi i}\ \frac{\mathrm{e}^{-u n}}{u^{2n-2\Delta+D-2} (v{-}u)^{\Delta}}\left(1+\frac{(u{+}2) (v{-}u)}{4}+\frac{ (v{-}u)^2}{12}\right)^{\!n} \!\! .
\ee
Let us now change variables to 
\be 
t=(v-u)\sqrt{n}\ .
\ee
The integral then becomes
\begin{align} 
\beta_{n,n-\Delta}^{D} &\sim -\frac{1}{2}\oint_{u=0} \frac{\mathrm{d}u}{2\pi i}\oint_{t=0} \frac{\mathrm{d}t}{2\pi i}\ \frac{\mathrm{e}^{-u n}n^{\frac{\Delta-1}{2}}}{u^{2n-2\Delta{+}D{-}2} t^{\Delta}}\left(1+\frac{(u+2) t}{4\sqrt{n}}+\frac{t^2}{12n}\right)^n \\
&\sim -\frac{1}{2}\oint_{u=0} \frac{\mathrm{d}u}{2\pi i}\oint_{t=0} \frac{\mathrm{d}t}{2\pi i}\ \frac{\mathrm{e}^{-u n}n^{\frac{\Delta-1}{2}}}{u^{2n-2\Delta{+}D{-}2} t^{\Delta}}\exp\left(\frac{(u+2)\sqrt{n} t}{4}+\frac{t^2}{12}\right) \ ,
\end{align}
where we used that $(1+\frac{a}{n})^n \sim \mathrm{e}^{a n}$ in the second line. As a next step, we interchange the order of the two integrals and perform the $u$ integral. This leads to
\begin{multline} 
\beta_{n,n-\Delta}^{D} \sim \frac{n^{2n-\frac{3}{2}\Delta+D-\frac{7}{2}}}{2(2n-2\Delta{+}D{-}3)!}  \\
\times \oint_{t=0} \frac{\mathrm{d}t}{2\pi i\, t^{\Delta}} 
\left(1-\frac{1}{4\sqrt{n}}t\right)^{2n-2\Delta{+}D{-}3} \exp \left(\frac{1}{2}\sqrt{n} t+\frac{t^2}{12}\right)\ .
\end{multline}
We can now use the Stirling approximation of the factorial to simplify the prefactor. Since $n$ is large, we can also remove the constant terms in the exponent of the first factor of the remaining integral as follows
\be 
\beta_{n,n-\Delta}^{D} \sim \frac{2^{1{-}D{-}2n+2\Delta}\mathrm{e}^{2n}n^{\frac{\Delta-2}{2}} }{\sqrt{\pi}}\oint_{t=0} \frac{\mathrm{d}t}{2\pi i\, t^{\Delta}} 
\left(1-\frac{1}{4\sqrt{n}}t\right)^{2n} \exp \left(\frac{1}{2}\sqrt{n} t+\frac{t^2}{12}\right)\ .
\ee
Next, we use that
\be 
\left(1-\frac{1}{4\sqrt{n}}t\right)^{2n} \exp \left(\frac{1}{2}\sqrt{n} t\right)=\exp\left(-\frac{t^2}{16}\right)+\mathcal{O}\left(n^{-\frac{1}{2}}\right)\ .
\ee
The remaining $t$ integral is then
\be 
\oint_{t=0} \frac{\mathrm{d}t}{2\pi i\, t^{\Delta}} \exp\left(\frac{t^2}{48}\right)=\frac{1}{\left(\frac{\Delta-1}{2}\right)!} \left(\frac{1}{48}\right)^{\frac{\Delta-1}{2}}\ .
\ee
Recall that $\Delta=n-j$ is odd, so that this is non-vanishing.
Putting the pieces together leads then to the following asymptotics:
\begin{tcolorbox}
\be 
\text{fixed $\Delta=n-j$,\quad large $n$:}\qquad \beta_{n,n-\Delta}^{D} \sim  \frac{2^{3{-}D{-}2n}\mathrm{e}^{2n} }{\sqrt{\pi n}\left(\frac{\Delta-1}{2}\right)!}\left(\frac{n}{3}\right)^{\frac{\Delta-1}{2}}\ . \label{eq:Regge asymptotics}
\ee
\end{tcolorbox}\noindent
For the actual residue $B_{n,n-\Delta}^{D}$, this means
\be 
B_{n,n-\Delta}^{D} \sim \frac{3^{\frac{1-\Delta }{2}} 2^{\frac{3}{2}-2 n} e^n \Gamma \left(\frac{D-3}{2}\right) n^{\frac{\Delta-D +1}{2}}}{\sqrt{\pi } \Gamma \left(\frac{\Delta
   +1}{2}\right)}\ ,
\ee
which decays exponentially with rate $\frac{e}{4}$. 
In particular, the asymptotic values are again positive. We should also remark that the error of this approximation is actually of order $\mathcal{O}(n^{-1})$. From our derivation it seems that it is of order $ \mathcal{O}(n^{-\frac{1}{2}})$. However, this cannot be right, since for any fixed $\Delta$, one can just compute the exact formula by taking the residue in $v$ in eq.~\eqref{eq:Regge limit contour integral 1}. For example, we can compute explicitly for $\Delta=3$ that
\be 
\beta_{n,n-3}^{D} =\frac{n^{D+2n-8}\, (n{-}D{+}7)\, (2n{+}D{-}8)}{48\, (2n{+}D{-}7)!}
\ee 
and similarly for any other fixed $\Delta$, see Appendix~\ref{app:Regge} for more explicit formulas. This makes it clear that the square roots are just an artifact of our derivation.

The corresponding formula for the bosonic string can be derived analogously and gives
\begin{align} 
\text{fixed $\Delta=n-j$,\quad large $n$:}\qquad  \beta_{n,n-\Delta}^{\open,\bosonic} &\sim  \frac{2^{-1{-}D{-}2n}\mathrm{e}^{2(n+4)} }{\sqrt{\pi n}\left(\frac{\Delta-1}{2}\right)!}\left(\frac{n}{3}\right)^{\frac{\Delta-1}{2}}\ , \\
B_{n,n-\Delta}^{\open,\bosonic}&\sim \frac{ 2^{-2 n-\frac{5}{2}} e^{n+4} \Gamma \left(\frac{D{-}3}{2}\right)}{\sqrt{\pi } n^{\frac{1}{2} (D-4)}\left(\frac{\Delta -1}{2}\right)!} \left(\frac{n}{3}\right)^{\frac{\Delta-1}{2}}\ .
\end{align}
In the special case of $D=4$ and $\Delta=1$, this coincides with \cite{Maity:2021obe}.

\acknowledgments
We thank Aaron Hillman, Dalimil Maz\'a\v{c} and Pinaki Banerjee for useful discussions. 
N.A-H., L.E.\ and S.M.\ are partially funded by the grant  DE-SC0009988 from the U.S.\ Department of Energy.
L.E.\ acknowledges funding from the Marvin Goldberger fund.
S.M.\ gratefully acknowledges the funding provided by Frank and Peggy Taplin. Y.H. is supported by Taiwan Ministry of Science and Technology Grant No. 109-2112-M-002 -020 -MY3. 

\appendix
\section{Some positive functions} \label{app:positive function}
In this appendix, we show that certain functions (power series) have only positive Laurent coefficients when expanded around $z=0$. We call such functions positive functions.

\subsection{Simple functions}
We start with
\begin{align} 
f(z)&=\frac{1}{\log(1-z)}+\frac{1}{z}\\
&=\frac{1}{2}+\frac{z}{12}+\frac{z^2}{24}+\frac{19z^3}{720}+\frac{3z^4}{160}+\frac{863 z^5}{60480 }+ \frac{275 z^6}{24192}+\mathcal{O}(z^7)\ .
\end{align}
As one can see the first few Taylor coefficients are all non-negative. To prove that in fact all Taylor coefficients are positive, we remark that we can write
\be 
z f(z)=\int_0^1 \mathrm{d}s\ \left(1-(1-z)^s\right)\ .
\ee
Multiplying the function by $z$ does not change positivity and hence it suffices to show that all coefficients of $z f(z)$ are positive. In this representation one can easily compute the $m$-th derivative at $z=0$:
\be 
\partial_z^m \left(z f(z) \right) \Big|_{z=0}=\int_0^1 \mathrm{d}s\ s \prod_{k=1}^{m-1} (k-s) \ .
\ee
The integrand on the RHS is manifestly positive in the region $0<s<1$, which proves positivity of $f(z)$.

A related positive function is
\begin{align}
g(z)&=\frac{1}{(z-1)\log(1-z)}-\frac{1}{z} \\
&=\frac{1}{2}+\frac{5 z}{12}+\frac{3 z^2}{8}+\frac{251 z^3}{720}+\frac{95 z^4}{288}+\frac{19087 z^5}{60480}+\mathcal{O}(z^6)\ .
\end{align}
Positivity becomes manifest by writing
\be 
z g(z)=\int_0^1 \mathrm{d}s\ \left((1-z)^{s-1}-1\right)\ ,
\ee
so that
\be 
\partial_z^m \left(z g(z) \right) \Big|_{z=0}=\int_0^1 \mathrm{d}s\ \prod_{k=1}^{m} (k-s) >0\ .
\ee
From the positivity of this function, we can also easily deduce that
\be 
h_{\alpha}(z)=\frac{1}{1-z}\left(-\frac{\log(1-z)}{z} \right)^\alpha
\ee
is a positive function for any $\alpha \ge -2$. To see this, we first note that it is true for $\alpha=-2$, because
\be 
h_{\alpha=-2}(z)=z^2 f'(z)+1\ ,
\ee
which is positive. Integrating $g(z)$ implies that the function
\be 
\int^z \mathrm{d}y\ g(y) = \log \left(-\frac{\log(1-z)}{z} \right)
\ee
has only positive coefficients. This then in turns gives positivity of
\be 
\exp \left( \beta \log \left(-\frac{\log(1-z)}{z} \right) \right)=\left(-\frac{\log(1-z)}{z} \right)^\beta
\ee
for any $\beta \ge 0$. Since
\be 
h_\alpha(z)=h_{\alpha=-2}(z)\left(-\frac{\log(1-z)}{z} \right)^{2+\alpha}
\ee
is now the product of two positive functions for $\alpha\geq-2$, it is also positive.
\subsection{Generating function of the \texorpdfstring{$D=10$}{D=10}, \texorpdfstring{$j=0$}{j=0} coefficients}
Let us show using the same ideas that the regular coefficients of 
\be
f(z)=\frac{(1-z)^2 \log(1-z)^2+(9 z-6) \log(1-z)+12}{18 (1-z)^2 \log(1-z)^5}
\ee 
are positive. This is the function that appears in the unitarity for the superstring in $D=10$ for $j=$, see Section~\ref{sec:unitarity in D=10}. The argument for this is very similar to the argument above, but is less aesthetic. We write
\begin{align}
36z^5 f(z)&=\int_0^1\mathrm{d}u\ (1-z)^{u-2} \big(-u^4 z^4+u^3 \left(3 z^5+2 z^4-4 z^3\right)\nonumber\\
&\qquad\qquad\qquad\qquad+u^2 \left(-z^6-7 z^5+8 z^4+12 z^3-12 z^2\right)\nonumber\\
&\qquad\qquad\qquad\qquad+u\left(2 z^6+3 z^5-23 z^4+6 z^3+36 z^2-24 z\right)\nonumber\\
&\qquad\qquad\qquad\qquad-z^6+2
   z^5+13 z^4-28 z^3-10 z^2+48 z-24\big)\ .
\end{align}
In this form, we can again compute derivatives of our function at zero, which gives for $n \ge 0$
\begin{multline} 
\partial_z^{n+6} \left(36z^5 f(z)\right)=(n^2+8n+15)\int_0^1 \mathrm{d}u\ u \prod_{j=1}^{n+1} (j-u) \\
\times \big(n^4 u^2-n^3 \left(2 u^3-8 u^2+3 u+1\right)+n^2 \left(u^4-14 u^3+24 u^2-7 u-1\right)\\
+n \left(6 u^4-28 u^3+26 u^2-2\right)+4 u \left(2 u^3-4
   u^2+u+1\right)\big)\ .
\end{multline}
The integrand is no longer manifestly positive and we have to work a bit harder to show positivity from here. As a first step, we notice that at large $n$, the term $n^4 u^2$ will dominate over the other ones in the parenthesis. Hence we get an estimate from below by minimizing the coefficients of $n^3$, $n^2$, $n$ and $n^0$ over $u \in [0,1]$. We further round them down to get nicer expressions. Thus, we have to show positivity of the integral
\be 
\int_0^1 \mathrm{d}u\ u \prod_{j=1}^{n+1} (j-u) \left(n^4 u^2-2\left(\frac{3}{4}n^3+n^2+n\right) \right) \ .
\ee
Since $\frac{3}{4}n^3+n^2+n \le n^3$ for $n \ge 5$, it finally suffices to show positivity of the integral 
\be 
\int_0^1 \mathrm{d}u\ u \prod_{j=1}^{n+1} (j-u) \left(n u^2-2 \right)
\ee
for sufficiently large $n$. The first few orders can then be checked directly on the computer. Clearly, the integral is positive in the region $u \ge \frac{2}{\sqrt{n}}$. Thus, we need to show that
\be 
\int_{\sqrt{\frac{2}{n}}}^1 \mathrm{d}u\ u \prod_{j=1}^{n+1} (j-u) \left(n u^2-2 \right) \ge \int_{0}^{\sqrt{\frac{2}{n}}} \mathrm{d}u\ u \prod_{j=1}^{n+1} (j-u) \left(2-n u^2 \right)
\ee
for sufficiently large $n$. Let us bound the left-hand side from below as follows:
\begin{align} 
\int_{\sqrt{\frac{2}{n}}}^1 \mathrm{d}u\ u \prod_{j=1}^{n+1} (j-u) \left(n u^2-2 \right) &\ge \int_{\sqrt{\frac{2}{n}}}^1 \mathrm{d}u\ u (1-u) \prod_{j=2}^{n+1} (j-1) (n u^2-2) \\
&= n! \left(\frac{n}{20}-\frac{1}{3}+\frac{1}{n}-\frac{8\sqrt{2}}{15 n^{\frac{3}{2}}} \right) \\
&\ge \frac{(n+1)!}{30}
\end{align}
where the last inequality holds for $n \ge 30$. For the right-hand side, we have instead
\begin{align}
\int_{0}^{\sqrt{\frac{2}{n}}} \mathrm{d}u\ u \prod_{j=1}^{n+1} (j-u) \left(2-n u^2 \right) &\le \int_{0}^{\sqrt{\frac{2}{n}}} \mathrm{d}u\ u \prod_{j=1}^{n+1} j \left(2-n u^2 \right) \\
&=\frac{(n+1)!}{n}\ .
\end{align}
Thus for $n \ge 30$, the left-hand side becomes greater than the right-hand side and all Taylor coefficients with $n \ge 30$ have to be positive. It is then a simple matter to check on the computer that also the first 30 coefficients are positive.

\section{Single-variable generating function} \label{app:single variable generating function}
We can also write $\beta_{n,j}^{D}$ in terms of a single, more complicated contour integral. To keep formulas simpler, we will focus on the supersymmetric case with $a=0$, but similar manipulations can be done for the bosonic formula. Starting from \eqref{eq:double contour formula}, we first integrate by parts as follows
\begin{align}
\beta_{n,j}^{D}&=-\frac{1}{n}\oint_{u=0} \frac{\mathrm{d}u}{2\pi i}\oint_{v=0} \frac{\mathrm{d}v}{2\pi i}\ \frac{(v-u)^j}{(uv)^{j+\frac{D-2}{2}}} \left(\partial_u+\partial_v \right) \frac{1}{(\mathrm{e}^v-\mathrm{e}^u)^{n+2a}} \\
&=\frac{1}{n}\oint_{u=0} \frac{\mathrm{d}u}{2\pi i}\oint_{v=0} \frac{\mathrm{d}v}{2\pi i}\ \frac{\mathrm{e}^{-a(u+v)}}{(\mathrm{e}^v-\mathrm{e}^u)^{n}}\left(\partial_u+\partial_v \right)  \frac{(v-u)^j}{(uv)^{j+\frac{D-2}{2}}} \\
&=-\frac{j+\frac{D-2}{2}}{n}\oint_{u=0} \frac{\mathrm{d}u}{2\pi i}\oint_{v=0} \frac{\mathrm{d}v}{2\pi i}\ \frac{(v-u)^j(u+v)}{(uv)^{j+\frac{D}{2}}(\mathrm{e}^v-\mathrm{e}^u)^{n}} \ .
\end{align}
We next change variables to 
\be 
x=\frac{uv}{u-v}\ , \qquad t=\frac{v-u}{2}\ .
\ee
This transforms the integral to
\begin{align}
\beta_{n,j}^{D}&=\frac{2^{\frac{2-D}{2}}(-1)^{\frac{D}{2}}(2j{-}2{+}D)}{n} \oint_{t=0} \frac{\mathrm{d}t}{2\pi i \, t^{\frac{D-2}{2}}\left(1-\mathrm{e}^{2t}\right)^n} \oint_{x=0} \frac{\mathrm{d}x}{2\pi i\,  x^{j+\frac{D}{2}}} \ \mathrm{e}^{n t(1-\sqrt{1-\frac{2x}{t}})}.
\end{align}
We recognize the exponential as the generating function of the modified Bessel function of the second kind $K_{m-\frac{1}{2}}(nt)$ and we have \cite{Krall}
\be 
\oint_{x=0} \frac{\mathrm{d}x}{2\pi i\,  x^{m+1}} \ \mathrm{e}^{n t(1-\sqrt{1-\frac{2x}{t}})}=\frac{n^{m}}{m!} \mathrm{e}^{n t} \sqrt{\frac{2nt}{\pi}} K_{m-\frac{1}{2}}(nt)\ .
\ee
Hence we can write $\beta_{n,j}^{D}$ as
\begin{align} 
\beta_{n,j}^{D}=\frac{(-1)^{n+\frac{D}{2}} 2^{\frac{5}{2}-\frac{D}{2}-n}n^{j+\frac{D-3}{2}}}{\sqrt{\pi}\left(j+\frac{D-4}{2}\right)!} \oint_{t=0} \frac{\mathrm{d}t\, K_{j+\frac{D-3}{2}}(nt)}{2\pi i \, t^{\frac{D-3}{2}} \sinh(t)^n} \ .
\end{align}
Finally, one can also rewrite this in terms of the modified Bessel functions of the first kind $I_{\pm (j + \frac{D-3}{2})}(nt)$. We have in general the relation
\be 
K_{j+\frac{D-3}{2}}(nt)=-\frac{\pi}{2}\frac{I_{j+\frac{D-3}{2}}(nt)-I_{-j-\frac{D-3}{2}}(nt)}{\sin\left(\pi \left(j+\frac{D-3}{2}\right)\right)}
\ee
However, the two terms have opposite parity and using the constraint $n{+}j$ odd, the second term leads to an integrand of even parity which does not contribute to the residue. This then leads to
\be 
\beta_{n,j}^{D}=\frac{\sqrt{\pi}2^{\frac{3}{2}-\frac{D}{2}-n}n^{j+\frac{D-3}{2}}}{\left(j+\frac{D-4}{2}\right)!} \oint_{t=0} \frac{\mathrm{d}t\, I_{j+\frac{D-3}{2}}(nt)}{2\pi i \, t^{\frac{D-3}{2}} \sinh(t)^n} \ .
\ee
With some further knowledge about special functions, it is from here actually rather simple to land again on the initial expression for the coefficients $\beta_{n,j}^D$ in terms of decomposing the Pochhammer symbol into Gegenbauer polynomials. Using the integral representation (see, e.g. \cite[Ch.~VI, Sec.~6.15]{watson1995treatise})
\be 
I_\alpha(z)=\frac{2^{-\alpha}z^\alpha}{\sqrt{\pi}\Gamma(\alpha+\frac{1}{2})} \int_{-1}^1 \mathrm{d}x \ (1-x^2)^{\alpha-\frac{1}{2}} \mathrm{e}^{x z}\ ,
\ee
and inserting back the normalization constant $c_{n,j}^D$, we get
\begin{align}
    B_{n,j}^D&=\frac{(2j+D-3)n^{j-1}\Gamma\left(\frac{D-3}{2}\right)}{2^{j+n}\sqrt{\pi}\left(j+\frac{D-4}{2}\right)!} \int_{-1}^1 \mathrm{d}x\ (1-x^2)^{j+\frac{D-4}{2}}\oint_{t=0} \frac{\mathrm{d}t\, t^j \mathrm{e}^{x n t}}{2\pi i \, \sinh(t)^n}\\
    &=\frac{(2j+D-3) \Gamma\left(\frac{D-3}{2}\right)}{2^{j+n}n \sqrt{\pi}\left(j+\frac{D-4}{2}\right)!} \int_{-1}^1 \mathrm{d}x\ (1-x^2)^{j+\frac{D-4}{2}} \partial_x^j \oint_{t=0} \frac{\mathrm{d}t\, \mathrm{e}^{x n t}}{2\pi i \, \sinh(t)^n}\ .
\end{align}
We finally have upon substituting $z=1-\mathrm{e}^{-2t}$
\begin{align} 
\oint_{t=0} \frac{\mathrm{d}t\, \mathrm{e}^{x n t}}{2\pi i \, \sinh(t)^n}&=2^n \oint_{t=0}\mathrm{d}t\   \mathrm{e}^{(x-1)nt} (1-\mathrm{e}^{-2t})^{-n}\\
&=2^{n-1} \oint_{z=0} \mathrm{d}z\ z^{-n}(1-z)^{\frac{1-x}{2}t-1} \\
&=2^{n-1} n R_n^\I(x)\ ,
\end{align}
where the residue polynomial was defined in \eqref{eq:residue polynomial}. Thus, we recover
\be 
B_{n,j}^D=\frac{(2j+D-3) \Gamma\left(\frac{D-3}{2}\right)}{2^{j+1} \sqrt{\pi}\left(j+\frac{D-4}{2}\right)!} \int_{-1}^1 \mathrm{d}x\ (1-x^2)^{j+\frac{D-4}{2}} \partial_x^j R_n^\I(x)\ ,
\ee
which after integration by parts is the integral of the residue polynomial against the Gegenbauer polynomial $G_j^{(D)}(x)$ with the appropriate normalization. Hence following these steps backwards gives an alternative derivation of the double contour formula.

\section{\label{app:triple}Derivation of the triple-contour representation}

In this appendix we derive residue formulae for the Gegenbauer coefficients in the case of closed strings. For the sake of variety, we will follow a different route to that of Section~\ref{sec:proof}.

For completeness and in order to exhibit parallels to the open-string case, let us first revisit it.
Setting $\alpha'=1$, open-string tree-level amplitudes can be parametrized by the following worldsheet integral with $a \in \mathbb{Z}$,
\be\label{eq:Ast}
A^{\open}_a(s,t) := - s^{a-1} \int_{0}^{1} \d z\, z^{-s - 2a} (1-z)^{-t-1-a},
\ee
where $z$ is the cross-ratio of the positions of vertex operators. We have analytically extended the integrand as a holomorphic function of $z \in \mathbb{C}$ since contour deformations will be needed in the following steps.

The case $a=0$ corresponds to gluon scattering in type-I superstring (the polarization-dependent prefactors are positive and hence can be stripped-off without affecting the conclusions), while $a=1$ gives tachyon scattering in the bosonic string case. This means $m^2 = -a$. While the explicit form of the amplitudes will not be needed for our purposes, let us spell them out for completeness:
\be
A_a^{\open}(s,t) = \begin{dcases}
	\qquad \frac{\Gamma(-s)\Gamma(-t)}{\Gamma(1{-}s{-}t)} &\quad\text{for}\quad a=0\quad \text{(type-I)},\\
	-\frac{\Gamma(-1{-}s)\Gamma(-1{-}t)}{\Gamma(-2{-}s{-}t)} &\quad\text{for}\quad a=1\quad \text{(bosonic)}.
	\end{dcases}
\ee
This easily gives $\Res_{s=n} A_a^{\open} = \tfrac{1}{(n+a)!} \prod_{k=1+a}^{n-1+3a} (t + k)$. Note that for $a=0$, the massless exchange with $n=0$ is special because of the normalization in \eqref{eq:Ast}. Below we focus on $n >0$ since positivity of the massless pole can be easily checked by inspection.

We first convert the cut in the kinematic $s$-variable into a worldsheet residue. This can be achieved by compactifying the integration contour near $z=0$, in a way similar to the Hankel contour (see, e.g., \cite[App.~A.2.1]{Mizera:2019gea}):
\be
\int_0^{1} = \frac{1}{e^{-2\pi i s}-1} \oint_{|z|=\epsilon} + \int_{\epsilon}^1,
\ee
where $e^{-2\pi i s}$ is the monodromy around $z=0$ of the integrand at hand. The starting points of the two integrals on the right-hand side have to coincide and be on the same sheet. The result has the effect of manifesting all the $s$-channel poles, as well as making the integral converge for any value of $s$, which was not the case for \eqref{eq:Ast}. Performing the cut leaves us with
\be\label{eq:C4}
\oint_{s=n} \d s\, A^{\open}_a(s,t) = n^{a-1} \oint_{z=0} \d z\, z^{-n - 2a} (1-z)^{-t-1-a}.
\ee
Note that the integrand on the right-hand side is now single-valued around $z$ since $n+2a \in \mathbb{Z}$.

We can now plug in this result to the formula for the Gegenbauer coefficients $B_{n,j}^{\open,a}$, which expressed in terms of integrals over the Mandelstam invariants read
\be\label{eq:Bjn}
B_{n,j}^{\open,a} := (-1)^j c_{n,j}^m \int_{-n + 4m^2}^{0} \d t  \oint_{s=n} \frac{\d s}{2\pi i} \,A^{\open}_a(s,t)\, \partial_t^j  \big( {-}t (t{+}n {-} 4m^2) \big)^{j + \frac{D-4}{2}} .
\ee
The $t$-integral between $-n+4m^2$ and $0$ corresponds to integrating over all scattering angles in the $s$-channel (recall that $\cos \theta = 1 + 2t/(s - 4m^2)$), while the $s$-residue is the unitarity cut. The overall mass-dependent constant $c_{n,j}^m$ is a manifestly positive and given by
\be\label{eq:c constant}
c_{n,j}^m = \frac{2^{D-3}}{(n-4m^2)^{D+j-3}}\, \frac{(j{+}\tfrac{D-3}{2})\,\Gamma(\tfrac{D-3}{2})}{\sqrt{\pi}\, \Gamma(j{+}\tfrac{D-2}{2})} > 0.
\ee
The coefficients still depend on $D$, but we suppress it from the notation for clarity.
Substituting \eqref{eq:C4} commuting the $z$ and $t$ integrals gives
\be
B_{n,j}^{\open,a} = (-1)^j c_{n,j}^m\, n^{a-1} \oint_{z=0} \frac{\d z}{2\pi i} \frac{z^{-n-2a}}{(1-z)^{1+a}} \int_{-n-4a}^0 \frac{\d t}{(1-z)^{t}}\, \partial_t^j \big( {-}t (t{+}n {+} 4a) \big)^{j + \frac{D-4}{2}}\!\!.
\ee
Integrating by parts $j$ times in $t$ yields
\begin{align}
B_{n,j}^{\open,a} = c_{n,j}^m\, n^{a-1} \oint_{z=0} \frac{\d z}{2\pi i} &\frac{z^{-n-2a}}{(1-z)^{1+a}} (-\log(1{-}z))^j \\
&\times\underbrace{\int_{-n-4a}^0 \frac{\d t}{(1-z)^{t}}\, \big( {-}t (t{+}n {+} 4a) \big)^{j + \frac{D-4}{2}}}_{H_{n,j}^a(0) - H_{n,j}^a(-n-4a)},\nn
\end{align}
since all the boundary terms vanish. As a sanity check we see that using $-\log(1-z) = \sum_{k=1}^{\infty} z^k / k$ all the terms with too high spin, $j \geq n {+}2a$, vanish identically.

At this stage, let us consider the $t$-integral in the indefinite form on its own,
\be
H_{n,j}^{a}(t') = \int^{t'} \frac{\d t}{(1-z)^{t}}\, \big( {-}t (t{+}n {+} 4a) \big)^{j + \frac{D-4}{2}}.
\ee
We are interested in the combination $H_{n,j}^a(0) - H_{n,j}^a(-n{-}4a)$ and hence can ignore the integration constant. From now on we assume that $D$ is even. Notice that introducing $u = \log(1-z)$ we can rewrite the above integral as
\be
H_{n,j}^a(t') = e^{(n+4a)u}\, \partial_u^{j + \frac{D-4}{2}} \int^{t'} \d t \, e^{-(t+n+4a)u}\, t^{j + \frac{D-4}{2}}.
\ee
At this stage we recognize that the integrand is a total derivative,
\be
H_{n,j}^a(t') = -(j+\tfrac{D-4}{2})!\, e^{(n+4a)u}\, \partial_u^{j + \frac{D-4}{2}} \int^{t'} \d t \, \partial_t \left( \frac{e^{-(t+n+4a)u}}{u^{j+\frac{D-2}{2}}} \sum_{k=0}^{j+\frac{D-4}{2}} \frac{(tu)^k}{k!} \right).
\ee
Evaluated at $t'=0$, this is simply
\be\label{eq:P0}
H_{n,j}^a(0) = -(j+\tfrac{D-4}{2})!\, e^{(n+4a)u}\, \partial_u^{j + \frac{D-4}{2}} \left( \frac{e^{-(n+4a)u}}{u^{j+\frac{D-2}{2}}}\right).
\ee
In order to simplify the contribution from $H_{n,j}^a(-n{-}4a)$, we consider the change of variables $z \to z/(z-1)$ together with $t \to -t{-}n{-}4a$, giving
\begin{align}
&\oint_{z=0} \frac{\d z}{2\pi i} \frac{z^{-n-2a}}{(1-z)^{1+a}} (- \log(1{-}z))^j H_{n,j}^a(-n{-}4a) \nn\\
&\quad =(-1)^{n+j} \oint_{z=0} \frac{\d z}{2\pi i} \frac{z^{-n-2a}}{(1-z)^{1+a}} (- \log(1{-}z))^j H_{n,j}^a(0),
\end{align}
which for even $n{+}j$ cancels the $H_{n,j}^a(0)$ term and for odd $n{+}j$ it doubles it. We therefore have
\be
B_{n,j}^{\open,a} = 0 \quad\text{for}\quad n+j \in 2\mathbb{Z},
\ee
and from now on we focus on the case $n{+}j \in 2\mathbb{Z}+1$. Using \eqref{eq:P0} we have
\be
B_{n,j}^{\open,a} = 2(-1)^j c_{n,j}^m\, n^{a-1} (j+\tfrac{D-4}{2})! \oint_{u=0} \frac{\d u}{2\pi i} \frac{u^j\, e^{u(n+3a)}}{(1-e^u)^{n+2a}}\, \partial_u^{j+\frac{D-4}{2}} \left( \frac{e^{-u(n+4a)}}{u^{j+\frac{D-2}{2}}} \right).
\ee
Recall that this expression is a result of localizing in the forward limit ($t=0$) and on the cut ($s=n$), which in turn localized on the worldsheet OPE ($z=0$ or equivalently $u=0$).

In the final step we use integration by parts $j+\tfrac{D-4}{2}$ times in $u$, followed by rewriting the derivative as a residue integral using the identity
\be\label{eq:residue-identity}
\partial_u^{J} f(u) = (-1)^{J} \oint_{v=0} \frac{\d v}{2\pi i} \frac{J!}{v^{J + 1}} f(u-v)
\ee
with $J = j + \frac{D-4}{2}$. This results in
\be
B_{n,j}^{\open,a} = 2 c_{n,j}^m\, n^{a-1} [\Gamma(j{+}\tfrac{D-2}{2})]^2 \oint_{u=0} \frac{\d u}{2\pi i} \oint_{v=0} \frac{\d v}{2\pi i} \frac{(v-u)^{j} e^{-(u+v)a}}{(u v)^{j+\frac{D-2}{2}} (e^v - e^u)^{n+2a}}.
\ee
where the constant $c_{n,j}^m$ was given in \eqref{eq:c constant}. This yields \eqref{eq:B to beta} and \eqref{eq:B bosonic} for $a=0$ and $1$ respectively. Note the order of taking residues: exchanging $u \leftrightarrow v$ picks up a factor of $(-1)^{j+n} = -1$, so the two residues do not commute, which can be seen alternatively from the fact that the integrand has a pole at $u=v$.

Notice that the double-contour formula is not well-defined for odd $D$, since the integrand is not single-valued around $v=0$. Nevertheless, let us observe that for even $D$ one can use the commutator of residues familiar from two-dimensional CFT (see, e.g., \cite[Sec.~6.1.2]{DiFrancesco}):
\be
\oint_{v=0} \oint_{u=0} - \oint_{u=0} \oint_{v=0} = \oint_{u=0} \oint_{v=u}
\ee
together with the aforementioned anti-symmetry under $u \leftrightarrow v$ to find an alternative expression for the partial-wave expansion coefficients
\be\label{eq:B-alternative}
B_{n,j}^{\open,a} = - c_{n,j}^m\, n^{a-1} [\Gamma(j+\tfrac{D-2}{2})]^2 \oint_{u=0} \frac{\d u}{2\pi i} \oint_{v=u} \frac{\d v}{2\pi i} \frac{(v-u)^{j} e^{-(u+v)a}}{(u v)^{j+\frac{D-2}{2}} (e^v - e^u)^{n+2a}}
\ee
for odd $n{+}j$. At this stage notice that after performing the $v$ residue, non-analyticity in $u$ can only come through a factor of the form $(u^2)^{-D/2}$ for odd $D$. We found that in such cases the correct choice of branch is $-u^{-D}$. In the following sections we only consider the cases with $D \in 2\mathbb{Z}$.

It is time to move on to the closed-string case \eqref{eq:A-closed} with the explicit expressions spelled out in \eqref{eq:closed-cases}. Our starting point is compactifying the worldsheet integration contour close to $z=0$. For the integrand at hand, locally we have
\begin{multline}\label{eq:C1}
A^{\closed}_{a,b}(s,t) = - \frac{1}{2 \pi i} \frac{s^{a+b-2}}{e^{-2\pi i s}-1} \\ 
\times \oint_{z=0} \d z\,  z^{-s-2a} (1-z)^{-t-1-a}\, \oint_{\tilde{z}=0} \d\tilde{z}\, \tilde{z}^{-s-2b} (1-\tilde{z})^{-t-1-b} + \dots\, ,
\end{multline}
where the ellipsis denotes non-singular terms around $s \in \mathbb{Z}$. The additional minus sign arose because we traded an anti-holomorphic residue for a holomorphic one by replacing $\bar{z} \to \tilde{z}$. Therefore, the cut at $s=n$ gives directly the formula \eqref{eq:closed-residue}. We take $n>0$ from now on, since $n=0$ can be checked by hand.

In computing the coefficient $B_{n,j}^{\closed,a,b}$, the $z$ and $\tilde{z}$ variables get mixed together by the $t$-integration:
\begin{align}
B_{n,j}^{\closed,a,b} = (-1)^j c_{n,j}^m\, n^{a+b-2} &\oint_{z=0} \frac{\d z}{2\pi i} \frac{z^{-n-2a}}{(1-z)^{1+a}} \oint_{\tilde{z}=0} \frac{\d \tilde{z}}{2\pi i} \frac{\tilde{z}^{-n-2b}}{(1-\tilde{z})^{1+b}} \nn\\
&\times \int_{-n+4m^2}^{0} \frac{\d t}{[(1-z)(1-\tilde{z})]^{t}}\, \partial_t^j \Big( {-}t(t{+}n{-}4m^2)\Big)^{j+\frac{D-4}{2}}.
\end{align}
Recall that $m^2 = -\min(a,b)$. Performing integration by parts $j$ times in $t$ gives
\begin{align}
B_{n,j}^{\closed,a,b} = c_{n,j}^m\, n^{a+b-2}\! &\oint_{z=0} \frac{\d z}{2\pi i} \frac{z^{-n-2a}}{(1{-}z)^{1+a}} \oint_{\tilde{z}=0} \frac{\d \tilde{z}}{2\pi i} \frac{\tilde{z}^{-n-2b}}{(1{-}\tilde{z})^{1+b}} \Big({-} \log(1{-}z) {-} \log(1{-}\tilde{z}) \Big)^{j} \nn\\
&\times \underbrace{\int_{-n+4m^2}^{0} \frac{\d t}{[(1-z)(1-\tilde{z})]^{t}} \Big( {-}t(t{+}n{-}4m^2)\Big)^{j+\frac{D-4}{2}}}_{Q_{n,j}^{a,b}(0) \,-\, Q_{n,j}^{a,b}(-n+4m^2)},
\end{align}
since boundary terms vanish. At this stage we are interested in evaluating $Q_{n,j}^{a,b}(0) - Q_{n,j}^{a,b}(-n{+}4m^2)$, where
\be
Q_{n,j}^{a,b}(t') =  \int^{t'} \frac{\d t}{[(1-z)(1-\tilde{z})]^{t}} \Big( {-}t(t{+}n{-}4m^2)\Big)^{j+\frac{D-4}{2}}.
\ee
Repeating steps identical to those above, for $D \in 2\mathbb{Z}$ we obtain
\be
Q_{n,j}^{a,b}(0) = -(j+\tfrac{D-4}{2})! \, e^{(n-4m^2)(u+\tilde{u})}\, \partial_{u+\tilde{u}}^{j+\frac{D-4}{2}} \left( \frac{e^{-(n-4m^2)(u+\tilde{u})}}{(u+\tilde{u})^{j + \frac{D-2}{2}}} \right),
\ee
where we introduced $u = \log(1{-}z)$ and $\tilde{u} = \log(1{-}\tilde{z})$.
On the other hand, the contribution from $Q_{n,j}^{a,b}(-n{+}4m^2)$ can be evaluated by sending $z \to z/(z-1)$, $\tilde{z} \to \tilde{z}/(\tilde{z}-1)$, and $t \to -t{-}n{+}4m^2$:
\begin{align}
&\oint_{z=0} \frac{\d z}{2\pi i} \frac{z^{-n-2a}}{(1-z)^{1+a}} \oint_{\tilde{z}=0} \frac{\d \tilde{z}}{2\pi i} \frac{\tilde{z}^{-n-2b}}{(1-\tilde{z})^{1+b}} \Big(- \log(1{-}z) - \log(1{-}\tilde{z}) \Big)^{j} Q_{n,j}^{a,b}(-n{+}4m^2) \nn \\
&= (-1)^{j+1} \oint_{z=0} \frac{\d z}{2\pi i} \frac{z^{-n-2a}}{(1-z)^{1+a}} \oint_{\tilde{z}=0} \frac{\d \tilde{z}}{2\pi i} \frac{\tilde{z}^{-n-2b}}{(1-\tilde{z})^{1+b}} \Big(- \log(1{-}z) - \log(1{-}\tilde{z}) \Big)^{j} \nn \\
&\qquad\qquad\qquad\qquad\qquad\qquad\qquad\times (1-z)^{4(a+m^2)} (1-\tilde{z})^{4(b+m^2)} Q_{n,j}^{a,b}(0).
\end{align}
In the final line, the exponents of $(1-z)$ and $(1-\tilde{z})$ vanish for type-II and bosonic string theories (where $a = b = -m^2$), but not in the heterotic case. Therefore, in the former cases $B_{n,j}^{\closed,a,b}$ is zero for odd $j$, while no simplification happens for the heterotic string. In order keep the notation brief, let us introduce
\be
g_{a,b} := 1 + (-1)^j e^{4 [u (a+m^2) + \tilde{u}(b+m^2)]} =  \begin{dcases}
	2 \delta_{j\in 2\mathbb{Z}}  &\text{for}\quad a=b \quad \text{(type-II, bosonic)},\\
	1 + (-1)^j e^{4\tilde{u}}  &\text{for}\quad a \neq b \quad \text{(heterotic)},
\end{dcases}
\ee
labelled according to \eqref{eq:closed-cases}.
To summarize, in terms of $u$ and $\tilde{u}$, we have
\begin{align}
B_{n,j}^{\closed,a,b} = (-1)^{j+1} &c_{n,j}^m\, n^{a+b-2}\, (j{+}\tfrac{D-4}{2})! \oint_{u=0} \frac{\d u}{2\pi i} \frac{e^{u(n-a)}}{(1-e^u)^{n+2a}} \oint_{\tilde{u}=0} \frac{\d \tilde{u}}{2\pi i} \frac{e^{\tilde{u}(n-b)}}{(1-e^{\tilde{u}})^{n+2b}}  \nn\\
&\quad\times g_{a,b}\,(u+\tilde{u})^{j}\, e^{-4(u+\tilde{u})m^2} \partial_{u+\tilde{u}}^{j+\frac{D-4}{2}} \left( \frac{e^{-(n-4m^2)(u+\tilde{u})}}{(u+\tilde{u})^{j + \frac{D-2}{2}}} \right).\label{eq:C8}
\end{align}
At stage we can use the identity \eqref{eq:residue-identity} to rewrite the $(u{+}\tilde{u})$-derivative as a residue, giving
\begin{align}
B_{n,j}^{\closed,a,b} = (-1)^{j} c_{n,j}^m\, n^{a+b-2}\, &[\Gamma(j{+}\tfrac{D-2}{2})]^2 \oint_{u=0} \frac{\d u}{2\pi i} \oint_{\tilde{u}=0} \frac{\d \tilde{u}}{2\pi i} \oint_{v=0} \frac{\d v}{2\pi i} \nn\\
& \times \frac{g_{a,b}\, e^{-u a - \tilde{u}b + v(n-4m^2)} (u + \tilde{u})^j}{(1-e^u)^{n+2a} (1-e^{\tilde{u}})^{n+2b} [v(v-u-\tilde{u})]^{j+\frac{D-2}{2}}},
\end{align}
which gives \eqref{eq:closed-super} and \eqref{eq:closed-heterotic} in special cases. One might in principle massage this formula further, e.g., by a change of variables $(u,\tilde{u},v) \to (u {-} \tfrac{v}{2}, \tilde{u} {-} \tfrac{v}{2}, -v)$, however the expressions do not become significantly simpler.

Alternatively, we could have arrived at a quadruple-contour representation for $B_{n,j}^{\closed,a,b}$ as follows. After eq.~\eqref{eq:C8} we can split the derivatives in terms of $\partial_u$ and $\partial_{\tilde{u}}$:
\be
\partial_{u+\tilde{u}}^{j+\frac{D-4}{2}} = \left( \tfrac{1}{2} \partial_u + \tfrac{1}{2} \partial_{\tilde{u}} \right)^{j+\frac{D-4}{2}} = 2^{-j-\frac{D-4}{2}} \sum_{k=0}^{j+\frac{D-4}{2}} \binom{j{+}\frac{D-4}{2}}{k} \partial_{u}^k \, \partial_{\tilde{u}}^{j+\frac{D-4}{2}-k}.
\ee
Integrating by parts in $u$ and $\tilde{u}$ for each term in the above $k$ sum and writing the resulting derivatives as residues we find:
\begin{align}
B_{n,j}^{\closed,a,b} =&\, -2^{-j-\frac{D-4}{2}} c_{n,j}^m\, n^{a+b-2} [(j{+}\tfrac{D-4}{2})!]^2 \oint_{u=0} \frac{\d u}{2\pi i} \oint_{\tilde{u}=0} \frac{\d \tilde{u}}{2\pi i} \oint_{v=0} \frac{\d v}{2\pi i} \oint_{\tilde{v}=0} \frac{\d \tilde{v}}{2\pi i} \nn\\
&\times \frac{\hat{g}_{a,b}\, e^{a(3v-u) + b(3\tilde{v}-\tilde{u}) +4(v+\tilde{v}) m^2 } (v+\tilde{v}-u-\tilde{u})^j }{(e^v - e^u)^{n+2a} (e^{\tilde{v}} - e^{\tilde{u}})^{n+2b} (u+\tilde{u})^{j + \frac{D-2}{2}}} \sum_{k=0}^{j{+}\frac{D-4}{2}} \frac{1}{v^{k+1} \tilde{v}^{j+\frac{D-2}{2} - k}},
\end{align}
where $\hat{g}_{a,b} = g_{a,b}|_{\tilde{u} \to \tilde{u}-\tilde{v}}$. Finally, we use the simplification
\be
\sum_{k=0}^{j{+}\frac{D-4}{2}} \frac{1}{v^{k+1} \tilde{v}^{j+\frac{D-2}{2} - k}} = - \frac{v^{-j-\frac{D-2}{2}} - \tilde{v}^{-j-\frac{D-2}{2}}}{v - \tilde{v}}.
\ee
In summary, for the type-II ($a=b=0$) and bosonic ($a=b=1$) cases we have $B_{n,j}^{\closed,a,a} = 0$ for $j$ odd and for $j$ even:
\begin{align}
B_{n,j}^{\closed,a,a} &= 2^{-j-\frac{D-6}{2}} c_{n,j}^{m}\, n^{2(a-1)} [(j{+}\tfrac{D-4}{2})!]^2 \oint_{u=0} \frac{\d u}{2\pi i} \oint_{\tilde{u}=0} \frac{\d \tilde{u}}{2\pi i} \oint_{v=0} \frac{\d v}{2\pi i} \oint_{\tilde{v}=0} \frac{\d \tilde{v}}{2\pi i} \nn\\
&\quad\times\frac{e^{-a(u+\tilde{u}+v+\tilde{v})} (v+\tilde{v}-u-\tilde{u})^j }{(v - \tilde{v}) (u+\tilde{u})^{j + \frac{D-2}{2}} [(e^v - e^u)(e^{\tilde{v}} - e^{\tilde{u}})]^{n+2a} }\left( \frac{1}{v^{j+\frac{D-2}{2}}} - \frac{1}{\tilde{v}^{j+\frac{D-2}{2}}} \right).
\end{align}
Similarly, for heterotic string ($a=0$, $b=1$) we get
\begin{align}
B_{n,j}^{\closed,0,1} &= 2^{-j-\frac{D-4}{2}} \frac{c_{n,j}^m}{n} [(j{+}\tfrac{D-4}{2})!]^2 \oint_{u=0} \frac{\d u}{2\pi i} \oint_{\tilde{u}=0} \frac{\d \tilde{u}}{2\pi i} \oint_{v=0} \frac{\d v}{2\pi i} \oint_{\tilde{v}=0} \frac{\d \tilde{v}}{2\pi i} \nn\\
&\quad\times\frac{(e^{3\tilde{v}-\tilde{u}} + (-1)^j e^{3\tilde{u}-\tilde{v}}) (v+\tilde{v}-u-\tilde{u})^j }{(v - \tilde{v}) (u+\tilde{u})^{j + \frac{D-2}{2}} (e^v - e^u)^{n} (e^{\tilde{v}} - e^{\tilde{u}})^{n+2} }\left( \frac{1}{v^{j+\frac{D-2}{2}}} - \frac{1}{\tilde{v}^{j+\frac{D-2}{2}}} \right).
\end{align}

\section{\label{app:Regge}Some coefficients on the Regge trajectories}
In this appendix, we give explicit formulas for the $\beta_{n,n-\Delta}^{D}$ for the superstring (in the notation of Section~\ref{sec:Regge}) and for low values of $\Delta=n-j$.
\begin{subequations}
\begin{align}
    \Delta&=1:& \beta_{n,n-1}^{D}&=\frac{n^{2n+D-5}}{2 (2 n+D-5)!}\ , \\
    \Delta&=3:& \beta_{n,n-3}^{D}&=\frac{n^{2 n+D-8} (n-D+7) (2n+D-8)}{48 (2n+D-7)!}\ , \\
    \Delta&=5:& \beta_{n,n-5}^{D}&=\frac{n^{2 n+D-12} (2 n+D-10) (2 n+D-10) }{11520 (2 n+D-9)!} \big(5 D^2 n+2 D^2\\
    &&&\quad\ \ -10 D n^2 -92 D n-40 D+5 n^3+78 n^2+415 n+198\big)\ ,\nn \\
    \Delta&=7: & \beta_{n,n-7}^{D}&=\frac{n^{2 n+D-16} (2 n+D-12) (2 n+D-14) (2 n+D-16) }{5806080 (2 n+D-11)!} \\
    &&&\times \big(-35 D^3 n^2-42 D^3 n-16 D^3+105 D^2 n^3+1239 D^2 n^2+1542 D^2 n\nonumber\\
    &&&\quad\ +624 D^2-105 D n^4-2100 D n^3-14269 D n^2-18630
   D n-8048 D\nonumber\\
   &&&\quad\ +35 n^5+903 n^4+10211 n^3+53337 n^2+73994 n+34320\big)\ \nn.
\end{align} \label{eq:more regge limit coefficients}
\end{subequations}
\noindent
\!\!Higher orders can be easily computed, but the expressions become quickly unwieldy. The expressions can be easily translated to the actual coefficients $B_{n,n-\Delta}^{D}$ by multiplying with the constant $c_{n,n-\Delta}^D$.

\bibliographystyle{JHEP}
\bibliography{bib}
\end{document}